\documentclass[runningheads]{llncs}
\usepackage[T1]{fontenc}
\usepackage{graphicx}
\usepackage{algorithmic}
\usepackage{textcomp}
\usepackage{xcolor}
\usepackage{tcolorbox}
\usepackage{amsmath}
\usepackage{amssymb}
\usepackage{tikz}
\usetikzlibrary{positioning, arrows.meta}
\usepackage{float}
\usepackage[export]{adjustbox}
\usepackage{booktabs}
\graphicspath{{./Paper/images/}}
\usepackage{colortbl} 
\usepackage{array}
\usepackage{xspace}
\definecolor{lightgray}{gray}{0.9}
\usepackage{subcaption}
\usepackage{siunitx}
\usepackage{xurl}
\usepackage{hyperref}  
\usepackage{multirow}
\def\sys{\textsc{MalwarePT}\xspace}
\def\ssys{\textsc{MalPT}\xspace}

\newcommand{\mypar}[1]{\noindent\textbf{#1}\xspace}

\newcommand{\totalMalicious}{115,826\xspace}
\newcommand{\totalBenign}{31,010\xspace}
\newcommand{\totalSamples}{146,836\xspace}
\newcommand{\totalUnfilteredDataset}{567,639\xspace}
\newcommand{\totalPretrainingSamples}{50,796\xspace}

\newcommand{\totalFinetuningSamples}{93,030\xspace}

\newcommand{\totalApiCallPredictionFunctions}{131,288\xspace}
\newcommand{\totalApiCallPredictionTrainingFunctions}{65,502\xspace}
\newcommand{\totalApiCallPredictionValidationFunctions}{13,250\xspace}
\newcommand{\totalApiCallPredictionTestingFunctions}{52,536\xspace}
\newcommand{\totalApiCallPredictionAPIs}{1,280\xspace}

\newcommand{\totalFunctionalityClassificationSamples}{45,357}

\newcommand{\trainingFunctionSignatures}{84,918}
\newcommand{\validationFunctionSignatures}{17,071}
\newcommand{\testingFunctionSignatures}{334,690}
\newcommand{\testingNormalFunctionSignatures}{206,539}
\newcommand{\testingRuleShiftFunctionSignatures}{128,151}
\newcommand{\totalFunctionalities}{171\xspace}

\newcommand{\malwareDatasetOne}{31,010}

\newcommand{\pretrainingTokens}{15B\xspace}
\newcommand{\seqLength}{1,024\xspace}
\newcommand{\contentLength}{1,022\xspace}
\newcommand{\totalPretrainingWindows}{6.4\xspace}

\newtcolorbox{takeawaybox}{
  colback=teal!5,
  colframe=teal!55!black,
  boxrule=0.5pt,
  arc=1mm,
  left=4pt,
  right=4pt,
  top=4pt,
  bottom=4pt,
  title=Key Takeaway,
  fonttitle=\bfseries
}

\def\BibTeX{{\rm B\kern-.05em{\sc i\kern-.025em b}\kern-.08em
    T\kern-.1667em\lower.7ex\hbox{E}\kern-.125emX}}

\begin{document}
\pagestyle{plain}
\title{MalwarePT: A Binary-Level Foundation Model for Malware Analysis}

\author{
    Saastha Vasan\inst{1},
    Yuzhou Nie\inst{1},
    Kaie Chen\inst{1} \\
    Yigitcan Kaya\inst{1},
    Hojjat Aghakhani\inst{1},
    Roman Vasilenko\inst{2} \\
    Wenbo Guo\inst{1},
    Christopher Kruegel\inst{1},
    Giovanni Vigna\inst{1,3}
}
\authorrunning{S. Vasan et al.}
\institute{
    University of California, Santa Barbara \\
    \email{\{saastha, yuzhounie, kaiechen, yigitcan, henrygwb, vigna\}@ucsb.edu} \\
    \email{\{hojjat, chris\}@cs.ucsb.edu}
    \and
    Cisco Systems \\
    \email{rovasile@cisco.com}
    \and
    Broadcom
}
\maketitle
\begin{abstract}
Automated malware analysis increasingly relies on machine learning, yet most existing methods remain task-specific and depend on handcrafted features or narrowly scoped models.
Recent developments in binary-level foundation models suggest a path toward reusable program representations, but their application to malware analysis remains underexplored, and most still operate at byte-level tokenization, limiting their ability to capture multi-byte code patterns.

In this work, we introduce \sys{}, a binary-level foundation model for malware analysis built on a ModernBERT-style encoder and pretrained with masked language modeling on Windows PE code-section bytes.
We study whether a single pretrained encoder can transfer across malware-analysis tasks at different granularities, and how tokenization design affects that transfer.
We train a byte-pair encoding (BPE) tokenizer on code-section bytes to compress frequent multi-byte patterns within a fixed context budget.

We evaluate \sys{} on three downstream tasks spanning token-, function-, and document-level prediction: API call prediction, functionality classification, and malware (program) detection under temporal drift.
Our evaluation demonstrates that pretraining yields substantial gains for API call prediction and functionality classification, and that increasing the BPE vocabulary beyond the byte-level baseline improves performance, with the strongest overall tradeoff at a vocabulary size of 1,024 tokens.
In malware detection at $\text{FPR} \approx 0.001$, \sys{} outperforms the neural network baselines, and is complementary to feature-engineering models that rely on PE structure.
We also compare against existing binary foundation models and show that \sys{}'s design choices yield gains across all downstream tasks.
\end{abstract}

\section{Introduction}
\label{sec:introduction}
Malware analysis encompasses a range of tasks that operate at different levels of granularity, from detecting whether an executable is malicious, to classifying it into a malware family, to identifying the specific capabilities embedded in individual functions~\cite{fuyong2017malware,kruczkowski2014support,kalash2018malware,pascanu2015malware,vasan2020imcfn,sajid2021soda}. 
In practice, these tasks are frequently solved independently, with separate feature extraction pipelines, model architectures, and training procedures.
This fragmentation is costly: each task is trained on its own limited labeled dataset, and representations learned for one task rarely transfer to another.

Prior machine-learning approaches to malware analysis largely fall into two categories: 
Feature-engineering methods rely on handcrafted attributes such as file headers, import tables, byte histograms, and opcode statistics~\cite{anderson2018ember,shafiq2009pe,santos2013opcode,saxe2015deep,ahmadi2016novel}. 
These approaches can be effective, but the resulting representations are typically tailored to a single downstream objective.
Raw-byte neural models avoid manual feature design by operating directly on executable bytes~\cite{raff2018malware,raff2021classifying,vasan2020imcfn,kalash2018malware}, but they too are usually trained separately for each task and do not produce transferable representations across different analysis tasks.

Recent work on foundation models pretrained on binary code offers a path toward reusable program representations.
Transformer encoders pretrained on executable bytes have shown promise for reverse-engineering tasks such as function boundary detection, code-region identification, and binary similarity~\cite{pei2020xda,benkraouda2024you,koo2021semanticawarebinarycoderepresentation,ahn2022practical}. 
However, their application to malware analysis remains limited in two important respects. First, existing binary foundation models are typically pretrained on benign, compiler-optimized benchmark binaries~\cite{pei2020xda} or are designed for reverse-engineering objectives that differ from malware-analysis tasks, leaving it unclear whether their representations transfer to malicious code distributions.
Second, most existing models operate with byte-level tokenization, where every byte is an independent token in a fixed 256-symbol vocabulary. 
This granularity forces the model to process frequent multi-byte patterns — common instruction sequences, API-call stubs, function prologues — as long chains of individual tokens, consuming context budget and reducing the amount of executable code the model can attend to within a fixed sequence length.

In this work, we ask whether a single pretrained encoder over malware code bytes can transfer across malware-analysis tasks at different granularities, and how tokenization design affects that transfer. 
To answer these questions, we introduce \sys{}, a binary-level foundation model for malware analysis.
\sys{} is pretrained with masked language modeling~\cite{devlin2018bert} on code-section bytes extracted from malicious Windows PE files. 
It uses a byte-pair encoding (BPE) tokenizer~\cite{radford2019language} to represent frequent multi-byte patterns more compactly, and a ModernBERT-style bidirectional encoder~\cite{warner2025smarter} with hybrid local-global attention for efficient long-sequence processing. 
By pretraining on malicious code distributions rather than benign benchmarks, and by moving beyond byte-level tokenization, \sys{} addresses the two limitations identified above.

To support our study, we curate a dataset of \totalSamples{} unpacked Windows PE executable files from an initial pool of \totalUnfilteredDataset{} samples, yielding \totalMalicious{} malicious and \totalBenign{} benign binaries for pretraining and downstream evaluation.
We evaluate \sys{} on three downstream tasks spanning progressively broader uses of code context. 
At the token level, we introduce an API-call prediction task in which the model recovers a masked API-call from surrounding function bytes. 
At the function level, we evaluate functionality classification, where the model predicts a CAPA-derived behavior label from raw function bytes~\cite{capa}. 
At the document level, we evaluate malware detection under temporal drift, where the model must aggregate evidence across many token sequences covering an executable's code section while operating at a strict low false-positive rate. 
We also study vocabulary size across five BPE configurations and compare against prior binary foundation models, including XDA~\cite{pei2020xda}, and recent malware language models~\cite{kurlandski2026beyond}.

Three main findings emerge from our results. 
First, pretraining substantially improves both API-call prediction and functionality classification relative to training from scratch, demonstrating that the pretrained encoder transfers beyond its masked-language-modeling objective to downstream tasks at multiple granularities. 
Second, moving beyond byte-level tokenization improves downstream performance, with the best results obtained at a vocabulary size of 1,024 tokens. This suggests that tokenization granularity is an important design choice, but that larger vocabularies do not necessarily lead to better performance.
Third, malware detection under strict low-FPR temporal drift is the hardest setting: \sys{} outperforms all neural baselines we evaluate, including raw-byte models and feature-engineering neural networks, demonstrating that pretrained code-section representations capture discriminative signals that these architectures miss.

\noindent In summary, we make the following contributions:
\begin{itemize}
\item We introduce \sys{}, a binary-level foundation model for malware analysis pretrained with masked language modeling on raw code-section bytes from malicious Windows PE files. 
We show that a single pretrained encoder transfers across token-, function-, and document-level malware-analysis tasks.
\item We introduce a new API-call prediction task---a token-level task that complements existing function- and document-level tasks---that evaluates masked API recovery directly from raw function bytes, providing a new probe of how well pretrained binary representations capture local code patterns.
\item We provide a systematic empirical study of tokenization granularity, identifying 1,024-token BPE as the strongest overall vocabulary choice among five configurations.
\item We show that under strict low-FPR malware detection with temporal drift, \sys{} outperforms all evaluated neural baselines, including raw-byte and feature-engineering neural network models.
\end{itemize}

\section{Methodology}
\label{sec:approach}
\begin{figure*}
    \vspace*{-\baselineskip}
    \centering
    \includegraphics[clip, trim=0.1cm 7.15cm 3.73cm 0.0cm, width=\textwidth]{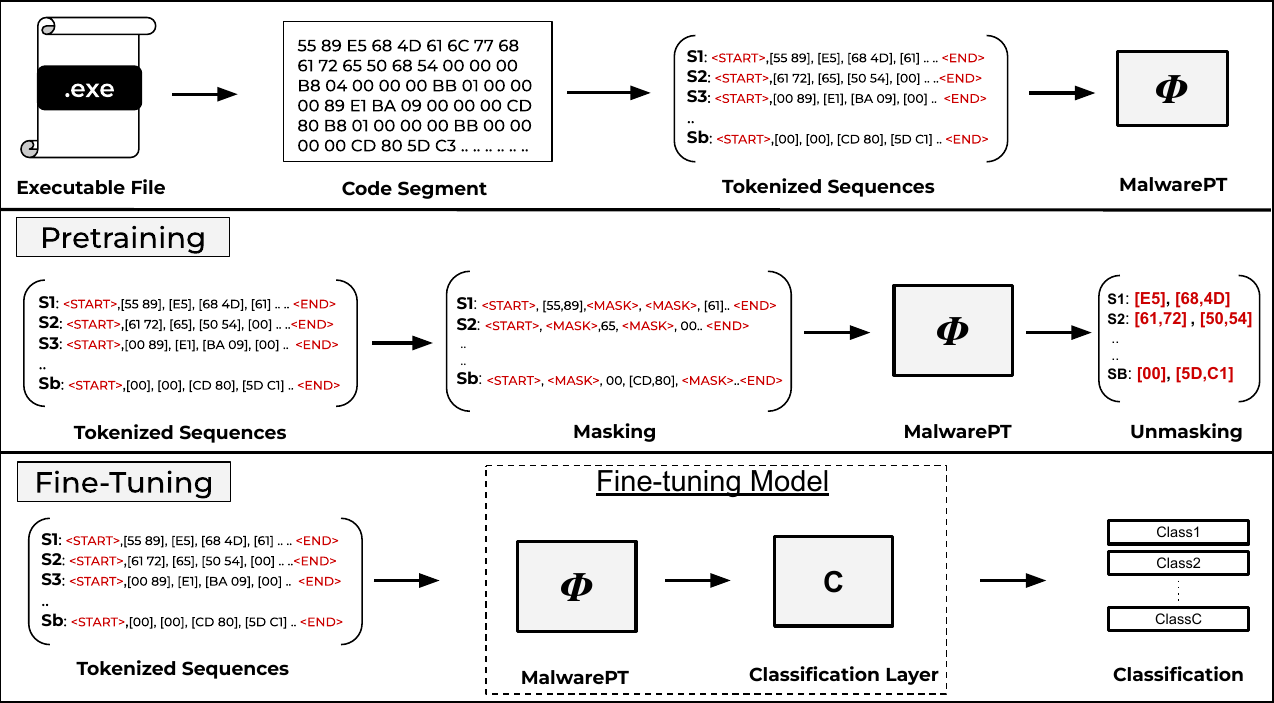}
    \caption{Overview of \sys{}. Raw bytes are extracted from the code section of each PE file, converted into atomic byte-level symbols, tokenized with a BPE vocabulary, and split into fixed-length sequences. These sequences are used to pretrain a ModernBERT-inspired bidirectional encoder with masked language modeling, after which the pretrained encoder is fine-tuned end-to-end with task-specific heads for downstream malware-analysis tasks.}
    \label{fig:system}
    \vspace{-0.2cm}
\end{figure*}

We introduce \sys{}, a foundation model for malware analysis trained on raw Windows PE code bytes. 
Our pipeline consists of three stages: extracting raw bytes from the executable code segment, converting those bytes into fixed-length token sequences using a BPE tokenizer, and pretraining a transformer encoder with masked language modeling before adapting it to downstream malware-analysis tasks with lightweight task-specific heads. Figure~\ref{fig:system} provides an overview.

\subsection{Tokenization}
Most prior binary transformer models represent each byte as an independent token, yielding a fixed vocabulary of 256 byte values. While simple, this representation is often too fine-grained for function-level and segment-level malware analysis: individual bytes carry limited contextual information, and byte-level tokenization produces long sequences that increase computational cost and make long-range pattern learning more difficult. To address this limitation, \sys{} uses an atomic Byte-Pair Encoding (BPE) tokenizer~\cite{radford2019language} trained directly on raw malware bytes.

Because BPE operates on symbol sequences rather than raw byte arrays, we first map each byte value (0x00--0xFF) to a unique Unicode character using a bijective encoding identical to the GPT-2 byte encoder~\cite{radford2019language}. This mapping is lossless and preserves a one-to-one correspondence between bytes and symbols, so every input remains exactly recoverable. BPE is then trained over this 256-symbol alphabet, allowing frequent multi-byte patterns to be merged into single tokens while retaining the original byte-level symbols as fallbacks. As a result, the tokenizer can represent recurring instruction-level patterns more compactly without introducing out-of-vocabulary failures.

Tokenizer training uses non-overlapping 512-byte windows extracted from the code sections of malicious PE binaries, a window size chosen to approximate the scale of individual functions and instruction sequences while remaining small enough to yield a large training corpus; corpus construction and filtering details are provided in Section~\ref{sec:tokenizer_training}. 
Each tokenizer additionally includes five special tokens: \emph{\textless s\textgreater}, \emph{\textless /s\textgreater}, \emph{\textless pad\textgreater}, \emph{\textless unk\textgreater}, and \emph{\textless mask\textgreater}.

After tokenization, each binary or function is split into fixed-length sequences of $N=\seqLength$ tokens, including the boundary tokens \emph{\textless s\textgreater} and \emph{\textless /s\textgreater}. If the final sequence is shorter than $N$, it is padded with \emph{\textless pad\textgreater}. This preprocessing produces uniform inputs for pretraining and fine-tuning while preserving the original byte content of the executable code segment.
For malware detection, this chunking is essential rather than incidental. 
The code section of a full executable often exceeds the encoder's context window by a large margin, so treating the entire binary as one token sequence would require aggressive truncation or a substantially different long-context design. 
We therefore keep the encoder input at a fixed length and represent each executable as an ordered list of token sequences during downstream document-level classification.

\subsection{Neural Network Architecture}
\vspace{-0.1cm}
\label{sec:neural_network_architecture}

\sys{} is a bidirectional transformer encoder, and its architecture is inspired by ModernBERT~\cite{warner2025smarter}, which modernizes the original BERT encoder~\cite{devlin2018bert} for improved efficiency and longer-context processing. 
Concretely, \sys{} retains the encoder-only, bidirectional formulation of BERT while adopting several modern architectural choices, including Rotary Position Embeddings~\cite{su2024roformer} (RoPE), RMSNorm-based pre-normalization, and a hybrid sparse attention pattern~\cite{beltagy2020longformer} with a globally connected sequence token.

\sys{} uses 12 transformer layers, 12 attention heads, hidden dimension $H=768$, and a feed-forward dimension of $3,072$, for a total of approximately $86M$ parameters.
We replace learned absolute positional embeddings with Rotary Position Embeddings (RoPE)~\cite{su2024roformer} and adopt pre-norm RMSNorm~\cite{zhang2019root} before each attention and feed-forward sub-layer for training stability.

\begin{figure}[t]
    \centering
    \resizebox{0.88\textwidth}{!}{\begin{tikzpicture}[x=0.095cm,y=0.095cm,font=\scriptsize]
    \def\n{20}
    \def\gap{34}
    \definecolor{attnfill}{RGB}{112,203,175}
    \definecolor{diagfill}{RGB}{29,145,83}
    \colorlet{emptyfill}{gray!6}
    \colorlet{gridline}{black!68}

    \begin{scope}[shift={(0,0)}]
        \foreach \i in {0,...,19} {
            \foreach \j in {0,...,19} {
                \fill[attnfill] (\j,-\i) rectangle ++(1,-1);
            }
            \fill[diagfill] (\i,-\i) rectangle ++(1,-1);
        }
        \draw[step=1,gridline,line width=0.14pt] (0,0) grid (\n,-\n);
        \draw[black,line width=0.5pt] (0,0) rectangle (\n,-\n);
        \node[anchor=north west,align=left] at (-1.5,-22.9) {(a) BERT/XDA\\Full attention};
    \end{scope}

    \begin{scope}[shift={(\gap,0)}]
        \fill[emptyfill] (0,0) rectangle (\n,-\n);
        \foreach \i in {0,...,19} {
            \foreach \j in {0,...,19} {
                \pgfmathtruncatemacro{\dist}{abs(\i-\j)}
                \ifnum\dist<3
                    \fill[attnfill] (\j,-\i) rectangle ++(1,-1);
                \fi
            }
            \fill[diagfill] (\i,-\i) rectangle ++(1,-1);
        }
        \draw[step=1,gridline,line width=0.14pt] (0,0) grid (\n,-\n);
        \draw[black,line width=0.5pt] (0,0) rectangle (\n,-\n);

        \begin{scope}[shift={(22,-2)}]
            \foreach \i in {0,...,5} {
                \foreach \j in {0,...,5} {
                    \fill[attnfill] (\j,-\i) rectangle ++(1,-1);
                }
                \fill[diagfill] (\i,-\i) rectangle ++(1,-1);
            }
            \draw[step=1,gridline,line width=0.11pt] (0,0) grid (6,-6);
            \draw[black,line width=0.35pt] (0,0) rectangle (6,-6);
            \node[font=\tiny,align=center] at (3,-7.9) {periodic\\full};
        \end{scope}

        \node[anchor=north west,align=left] at (-1.5,-22.9) {(b) ModernBERT\\Local + periodic full};
    \end{scope}

    \begin{scope}[shift={(2*\gap,0)}]
        \fill[emptyfill] (0,0) rectangle (\n,-\n);
        \foreach \i in {0,...,19} {
            \foreach \j in {0,...,19} {
                \pgfmathtruncatemacro{\dist}{abs(\i-\j)}
                \ifnum\dist<3
                    \fill[attnfill] (\j,-\i) rectangle ++(1,-1);
                \fi
            }
        }
        \foreach \j in {0,...,19} {
            \fill[attnfill] (\j,0) rectangle ++(1,-1);
        }
        \foreach \i in {0,...,19} {
            \fill[attnfill] (0,-\i) rectangle ++(1,-1);
            \fill[diagfill] (\i,-\i) rectangle ++(1,-1);
        }
        \draw[step=1,gridline,line width=0.14pt] (0,0) grid (\n,-\n);
        \draw[black,line width=0.5pt] (0,0) rectangle (\n,-\n);

        \begin{scope}[shift={(22,-2)}]
            \foreach \i in {0,...,5} {
                \foreach \j in {0,...,5} {
                    \fill[attnfill] (\j,-\i) rectangle ++(1,-1);
                }
                \fill[diagfill] (\i,-\i) rectangle ++(1,-1);
            }
            \draw[step=1,gridline,line width=0.11pt] (0,0) grid (6,-6);
            \draw[black,line width=0.35pt] (0,0) rectangle (6,-6);
            \node[font=\tiny,align=center] at (3,-7.9) {periodic\\full};
        \end{scope}

        \node[anchor=north west,align=left] at (-1.5,-22.9) {(c) ModernBERT +\\global \emph{\textless s\textgreater}};
    \end{scope}
\end{tikzpicture}}
    \caption{Attention patterns in byte-level binary encoders. BERT-style encoders such as XDA use full self-attention; ModernBERT-style encoders alternate local sliding-window layers with periodic full-attention layers; \sys{} additionally keeps \emph{\textless s\textgreater} globally connected in local-attention layers.}
    \label{fig:attention_patterns}
    \vspace{-0.25cm}
\end{figure}
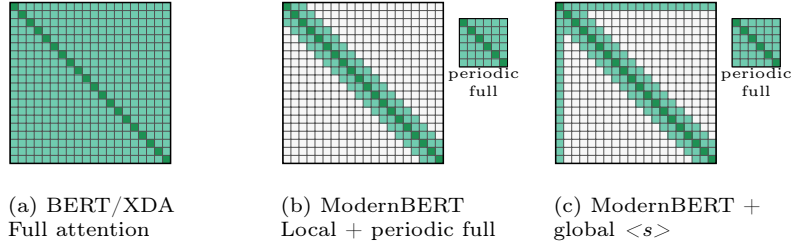

\sys{} uses a hybrid sparse attention pattern rather than full quadratic self-attention in every layer. 
Most layers apply sliding-window local attention with window size 128, which biases the model toward nearby instruction context and lowers computational cost. 
Every fourth layer uses full global attention to refresh long-range interactions across the entire sequence.
Critically, unlike standard sparse bidirectional encoders, the \emph{\textless s\textgreater} token at position 0 is globally connected in every layer, including local-attention layers, allowing it to aggregate information from the full sequence regardless of the attention pattern. 
This is essential for downstream use: the contextual embedding of \emph{\textless s\textgreater} serves as the sequence-level representation for both function-level classification and document-level aggregation, so it must have access to the complete sequence context at every layer. 
Figure~\ref{fig:attention_patterns} summarizes the attention patterns.
The model uses a standard masked-language-modeling prediction head consisting of LayerNorm, a linear projection, GELU, and a vocabulary projection tied to the input embedding matrix.

\subsection{Pretraining}
\sys{} is pretrained using masked language modeling (MLM)~\cite{devlin2018bert}. For each input sequence, 25\% of non-special tokens are selected for prediction. Among the selected tokens, 80\% are replaced with \emph{\textless mask\textgreater}, 10\% are replaced with a random token, and 10\% are left unchanged, following the standard BERT masking procedure. Masked positions are re-sampled dynamically during training in the style of RoBERTa~\cite{liu2019roberta}.
The MLM objective encourages the encoder to model inter-dependencies among tokens, including recurring local instruction patterns. 
We pretrain on \totalPretrainingSamples{} malicious executables. 
Dataset construction is detailed in Section~\ref{sec:dataset_generation}, and the full model architecture and pretraining hyperparameters for the main \sys{}-1024 configuration are listed in Appendix~\ref{app:model_architecture} (Table~\ref{tab:model_hyperparameters}). We optimize the model with cross-entropy loss over the masked positions.

\subsection{Adaptation to Downstream Tasks}
After pretraining, we discard the MLM head and fine-tune the encoder end-to-end for each downstream task by attaching a lightweight task-specific head. 
For API-call prediction, the model predicts masked API-call tokens within a function sequence using the surrounding context. 
For functionality classification, the contextual embedding of the \emph{\textless s\textgreater} token is passed to a classifier that predicts a CAPA-derived behavior label for an individual function. 

For malware detection, the downstream input is not a single sequence but an ordered list of fixed-length sequences covering the executable's code section (capped at 128 sequences, or $\sim$$128K$ total tokens). 
We encode each chunk independently and collect the contextual embedding of its \emph{\textless s\textgreater} token as a sequence-level representation.
We aggregate the \emph{\textless s\textgreater} representations across all chunks and pass them through RoPE-based multi-head self-attention followed by a linear classifier.
This lets the model capture ordering and interactions among chunks without forcing the pretrained model to process the full binary as one sequence, while keeping peak memory cost proportional to the fixed chunk length rather than the full executable.
The classification head was selected based on downstream performance; an ablation of alternative aggregation strategies is provided in Appendix~\ref{app:ablation_study}.
This design allows the same pretrained encoder to support token-level, sequence-level, and document-level malware-analysis tasks with minimal task-specific modification.

\section{Experimental Setup}
\label{sec:eval_setup}
\subsection{Dataset Generation}
\label{sec:dataset_generation}

Our dataset comprises \totalSamples{}  unpacked Windows PE executable files filtered from an initial collection of \totalUnfilteredDataset{} samples. Malicious samples were sourced from VirusTotal~\cite{virustotal} (87,754 retained; 31.7\% retention rate), Packware~\cite{aghakhani2020malware} (11,554 retained; 5.0\% retention rate), and DeepCapa~\cite{vasan2024deepcapa} (16,518 retained; 27.1\% retention rate). 
Benign samples were curated from Assemblage~\cite{liu2024assemblage}, DeepCapa, and PACE~\cite{Saha2024DRSM}. Overall, our filtering pipeline yields a final retention rate of 20.4\%.

\begin{figure*}[t]
    \centering
    \includegraphics[width=\textwidth]{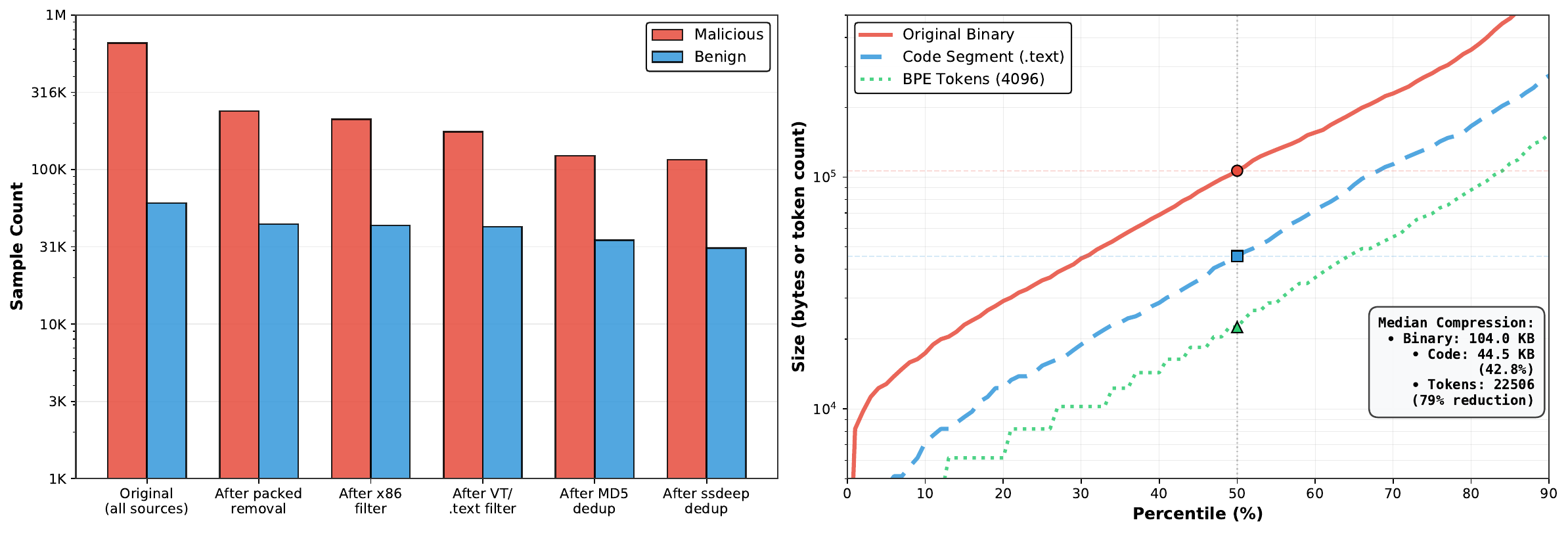}
    \caption{\textbf{Left:} Dataset filtering across six stages. 
    \textbf{Right:} Log scale showing original executable file size (red solid), code segments size (blue dashed), and BPE tokens (green dotted).
    BPE tokens reduce the original binary by approximately 79\%.
    }
    \label{fig:dataset_compression}
    \vspace{-0.3cm}
\end{figure*}

\mypar{Filtering Procedure.}
We apply a multi-stage filtering process (Figure~\ref{fig:dataset_compression}, left panel) to ensure label quality, eliminate train/test leakage, and preserve behavioral diversity. Following prior work~\cite{dambra2023decoding,kurlandski2026beyond}, we first remove packed samples using trID~\cite{TrID}, Detect It Easy (DIE)~\cite{DetectItEasy}, and Avast RetDec's YARA rules, eliminating over 420,000 malicious and 16,000 benign samples. We then retain only x86 32-bit PE files, ensuring a homogeneous binary format suitable for consistent tokenization and modeling.

For malicious samples, we discard those with fewer than 10 antivirus detections in their VirusTotal reports. Following recent work~\cite{kurlandski2026beyond}, we perform MD5-based deduplication on the code segment to prevent train/test leakage, removing malicious samples whose code section matches any benign sample. Finally, we apply code variant filtering by removing samples with $>95\%$ ssdeep~\cite{kornblum2006identifying} similarity on their code section and cap per-hash counts at two malicious and ten benign variants, ensuring balanced representation across variants. 
These filtering steps yield the final dataset of \totalSamples{} samples shown in Figure~\ref{fig:dataset_compression} (left).

\mypar{Discussion.}
While we detect and filter off-the-shelf packers (UPX, ASPack, PECompact, etc.), custom and proprietary packing techniques remain undetectable through static signatures.
We acknowledge that despite our rigorous filtering efforts, the dataset may still include custom-packed malware samples.
We discuss the effect of these samples on \sys{} for the segment-level downstream task in Section~\ref{sec:malware_detection}.
We further extend the discussion to outline ways to incorporate packed samples for instruction-level and function-level analysis in Section~\ref{sec:discussion}.

\mypar{Dataset Characteristics.}
To better understand the diversity of malware families in our dataset, we retrieved and analyzed the AV labels of all malicious executable files, using AVClass2~\cite{sebastian2020avclass2} to generate normalized malware family names.
AVClass2 failed to identify families for 7.04\% of the samples. 
Among the remaining 92.96\%, we identified 1,992 distinct malware families, representing a broad spectrum of relevant threats. 
The ten most prominent families---Virlock, Vobfus, Virut, Allaple, Hematite, Teslacrypt, Sality, Fareit, Locky, and Cerber---account for 46.9\% of all samples in our dataset.

\mypar{Temporal Splits.}
We employ a temporal evaluation strategy for malware detection to assess model robustness against distribution shift.
All \totalMalicious{} malicious samples are sorted chronologically by VirusTotal \texttt{first\_seen} timestamps to prevent train/test leakage and simulate realistic deployment scenarios.

We partition the malicious samples into a pretraining-only set ($M_2$) and a fine-tuning set ($M_1$).
The oldest 22,796 samples form $M_2$, reserved exclusively for pretraining.
The remaining \totalFinetuningSamples{} samples in $M_1$ are split into three equal temporal cohorts spanning distinct malware evolution periods: $D_1$ (2017--2018), $D_2$ (2018--2020), and $D_3$ (2020--2022), each containing \malwareDatasetOne{} samples and split 50/10/40 into train/validation/test.

The \totalBenign{} benign samples are split 50/10/40 and used exclusively for malware detection evaluation.
For malware detection, we evaluate temporal drift by testing on each temporal cohort separately (Section~\ref{sec:malware_detection}), pairing each cohort's test set with the benign test set to form balanced 50-50 malicious/benign examples.
Training and validation combine splits from all three cohorts with corresponding benign splits.
For all other downstream tasks (API-call prediction and functionality classification), we use only malicious samples, combining train/validation/test splits across $D_1$, $D_2$, and $D_3$.

\subsection{Environment Setup}
Our base system for this research comprises an Ubuntu 20.04 machine, with 1.5 TB of RAM, 2$\times$AMD EPYC 9554 64-core processors, and 8$\times$ NVIDIA L40S GPUs.
We use the PyTorch~\cite{Pytorch} library to implement our neural network model.
During the fine-tuning phase, for the functionality classification task, we extract the raw bytes of all the functions from an executable file using the IDA Pro disassembler~\cite{ida_pro}.


\mypar{Pretraining Data Preparation.}
We pretrain on all malicious samples in $M_2$ plus 28,000 samples from the training split of $M_1$, yielding approximately \totalPretrainingWindows{} million sequences (25~GB total).
Each binary's code section is tokenized and split into $N=\seqLength$-token sequences, with a Shannon entropy filter ($\leq 7.0$) applied to exclude packed content and retain clean samples.

\mypar{Tokenizer Training.}
\label{sec:tokenizer_training}
We train BPE tokenizers (vocab sizes 512, 1,024, 2,048, 4,096) on code section bytes from 10,000 samples chosen randomly from the pretraining corpus.
Bytes are split into 512-byte non-overlapping windows, mapped to Unicode via the bijective encoding (Section~\ref{sec:neural_network_architecture}), and fed to the BPE trainer.

BPE tokenization achieves substantial sequence compression. The code section comprises approximately 42\% of the original binary (median). With 4,096-token BPE, the code representation is further compressed to approximately 24\% of the original binary, yielding a 4.7$\times$ overall compression ratio (median). This compression allows \sys{} to maintain a fixed context budget while capturing substantially longer code sequences within individual tokens, reducing chunking overhead and enabling the model to learn longer-range dependencies.

\mypar{Pretraining Setup.}
\sys{} uses the ModernBert~\cite{warner2025smarter} architecture described in Section~\ref{sec:neural_network_architecture}, with an embedding dimension $H$ of 768.
We pretrain \sys{} using four vocabulary sizes: 512, 1,024, 2,048, and 4,096 tokens.
All models are pretrained for a budget of \pretrainingTokens{} tokens using an effective batch size of 128 sequences (64 per GPU $\times$ 2 gradient accumulation steps).
The learning rate is set to $10^{-4}$ with a 1\% linear warmup followed by polynomial decay.
We use BF16 mixed-precision training for efficiency.
Validation is performed every 2,500 optimizer steps, and checkpoints are saved every 5,000 steps.
Full model architecture and pretraining hyperparameters for the main \sys{}-1024 configuration are listed in Appendix~\ref{app:model_architecture} (Table~\ref{tab:model_hyperparameters}).

\mypar{Non-Pretrained Variant.}
To isolate the contribution of pretraining in our downstream evaluations, we compare \sys{} against a non-pretrained variant called \sys{}-N.
\sys{}-N uses the same model architecture as \sys{} but is initialized with random weights and trained from scratch on each downstream task, rather than loaded from the pretrained checkpoint.
Any performance gap between \sys{} and \sys{}-N therefore reflects the effect of pretraining independent of architecture choices.
Full fine-tuning hyperparameters for both variants are listed in Appendix~\ref{app:finetuning_setup}.

\section{Experimental Evaluation}
\label{sec:evaluation}

Our evaluation asks whether malware-specific pretraining over raw executable code yields transferable representations, and which design choices are responsible for that transfer.
We structure the evaluation around three downstream tasks with different granularities: token-level API-call prediction (Section~\ref{ssec:api_call_eval}), function-level functionality classification (Section~\ref{sec:eval_functionality_classification}), and document-level malware detection under temporal drift (Section~\ref{sec:malware_detection}).
We then compare \sys{} against prior binary foundation models (Section~\ref{sec:eval_comparison_sota}) and ablate vocabulary size to study tokenization granularity (Section~\ref{sec:eval_vocab_size}).
Supplementary experiments (fine-tuning architecture ablation, pretraining data domain) are reported in the Appendix.

\begin{table*}[t]
\centering
\caption{API-call prediction performance. For brevity, we refer to \sys{} as \ssys{}.}
\label{tab:api_call_prediction}
\resizebox{\textwidth}{!}{%
\begin{tabular}{|l *{4}{S[table-format=2.2]}|}
\toprule
\textbf{Model} & {\textbf{W.F1 Score (\%)}} & {\textbf{Top-1 Accuracy (\%)}} & {\textbf{Top-3 Accuracy (\%)}} & {\textbf{Top-5 Accuracy (\%)}} \\
\midrule

\rowcolor{gray!20}
\ssys{}-N 1024  & 64.54 & 64.33 & 72.24 & 74.72 \\
\midrule

\ssys{} 1024    & \textbf{86.55} & \textbf{87.83} & \textbf{92.17} & \textbf{93.01} \\
\bottomrule
\end{tabular}%
}
\end{table*}

\subsection{API-call Prediction}
\label{ssec:api_call_eval}
The experiments presented in this section measure \sys{}'s ability to resolve API-call names within a function.
We design these experiments as a masked token prediction task: given a tokenized byte sequence of a function where an API-call invocation is replaced with a \emph{\textless MASK\textgreater} token, \sys{} must correctly predict the masked API-call name.
Success in this task depends on both the unique patterns of individual API-calls and the contextual dependencies between calls and their surrounding bytes (call context).

The real-world application of the API-call prediction task is inferring the identity of indirect or obfuscated API-calls where the symbolic name is absent from the static disassembly.
Such API-calls make manual analysis more difficult and often degrade the effectiveness of automated static analysis tools.
Our evaluation serves two main goals: (1) to assess how effectively \sys{} uses contextual relationships between tokens, as reflected in API prediction accuracy; and (2) to evaluate the effectiveness of our pretraining approach by comparing the performance of fine-tuned \sys{} with \sys{}-N.

\mypar{Dataset and Task Setup.}
API-call prediction dataset comprises \totalApiCallPredictionFunctions{} functions that include calls to \totalApiCallPredictionAPIs{} unique Windows API functions.
We disassembled each malicious executable in our fine-tuning dataset to obtain the function's raw byte sequence and the offsets and names of all API-call invocations within it.
To balance the dataset, we filtered API-calls based on their overall frequency: we removed functions containing API-calls that occurred fewer than 100 times throughout the dataset.
Conversely, for API-call types occurring more than 100 times, we randomly select 100 functions for each type.
To enable direct prediction of API-calls as tokens, we extended \sys{}'s pretrained vocabulary by incorporating \totalApiCallPredictionAPIs new API-call tokens (see Section~\ref{sec:neural_network_architecture}.)
To prepare input sequences for the model, we provide a tokenized function sequence where specific API-call invocations are masked.
We partitioned the dataset into training, validation, and test sets in a 50\%, 10\%, and 40\% split, consisting of \totalApiCallPredictionTrainingFunctions, \totalApiCallPredictionValidationFunctions, and \totalApiCallPredictionTestingFunctions functions, respectively.
The detailed training setup for \sys{} and \sys{}-N for API-call prediction is explained in Appendix~\ref{app:finetuning_setup}.

\mypar{Results.}
Table~\ref{tab:api_call_prediction} summarizes the performance of \sys{} models for API-call prediction.
Performance is measured using weighted F1 Score and Top-$N$ accuracy ($N \in \{1, 3, 5\}$), which considers a model prediction correct if the true label is contained among the $N$ most probable predicted classes.

The results highlight the effect of pretraining.
Our \sys{}-1024 model achieves a weighted F1 Score of 86.55\% and a Top-1 Accuracy of 87.83\%, outperforming \sys{}-N 1024, which reaches only 64.54\% Weighted F1 Score and 64.33\% Top-1 Accuracy.
This clear advantage, representing an absolute improvement of roughly 22\% in weighted F1 Score, demonstrates that pretraining provides representations that transfer effectively to downstream prediction.

Notably, the Top-5 Accuracy reaches 93.01\%, suggesting that even when the model misses the exact call, the correct answer is highly likely to be within its top predictions.
This indicates that our BPE tokenization enables richer token representations, allowing the model to use specific instruction sequences or parameter patterns associated with API-calls more effectively.

\begin{takeawaybox}
Malware-specific pretraining substantially improves masked API-call recovery over training from scratch, indicating that pretrained binary-code representations transfer to token-level prediction.
\end{takeawaybox}

\subsection{Functionality Classification}
\label{sec:eval_functionality_classification}
Functionality classification of malware aims to provide analysts with insight into function-level behaviors. 
For example, malware may connect to C\&C server, perform file-system read and write operations, encrypt files, and modify registry keys to establish persistence. 
Identifying these operations through static analysis can help analysts map observed behaviors to MITRE ATT\&CK~\cite{mitre_attack} tactics, techniques, and procedures, and support the development of mitigation strategies.

Our objective in this experiment is to evaluate whether \sys{} can predict function-level behavioral labels from raw bytes. 
To obtain ground-truth functionality labels, we use CAPA~\cite{capa}.
CAPA identifies functionality using expert-crafted rules. 
To capture variations in how a behavior is implemented, these rules may encode multiple alternative feature patterns.
In our evaluation, we utilize CAPA not as a baseline, but as a high-fidelity oracle to generate functionality labels.
By fine-tuning \sys{} to predict these labels, we evaluate whether our architecture can map function byte sequences to CAPA-derived behavior categories without relying on manually-written rules at inference time.

\mypar{Dataset.}
To generate ground-truth labels, we run CAPA on all malicious executables in our fine-tuning dataset.
We use CAPA because its rules map directly to human-interpretable concepts, and its reports localize each detected behavior to a specific memory address, letting us tie functionalities to individual functions.
To keep the labels clean, we retain only functions where CAPA detects exactly one unambiguous functionality; this removes label noise and isolates distinct behaviors.
For each retained function, we use a disassembler to extract the raw byte sequence at the CAPA-reported address and assign the corresponding functionality label.
The resulting dataset contains \totalFunctionalities{} unique functionalities across \totalFunctionalityClassificationSamples{} malware samples.

\mypar{Train, Validation, and Test Splits.}
We partition the functions into training, validation, and test splits, and then apply three construction steps on the training and validation sides.
First, for every functionality that has two or more distinct CAPA rules, we hold out 50\% of its rules and discard the training and validation entries labeled by those held-out rules.
Second, we drop any functionality that has fewer than 50 remaining training instances.
Third, we cap each remaining functionality at 1{,}000 training instances by random subsampling so that frequent functionalities do not dominate.
The test split is left untouched and uncapped, and we partition it into two evaluation sets according to whether each test function's labeling rule was retained in training: a \emph{normal} set (rule seen in training) and a \emph{rule-shift} set (rule held out).
This procedure yields \trainingFunctionSignatures{} training, \validationFunctionSignatures{} validation, and \testingFunctionSignatures{} test functions, with the test split further divided into \testingNormalFunctionSignatures{} normal and \testingRuleShiftFunctionSignatures{} rule-shift entries.
We cap only the training split so that rare classes and rule-shift entries in the test split retain enough support for reliable per-class measurement.
Of the \totalFunctionalities{} functionality classes, 98 have non-zero support in the rule-shift split; the remaining 73 classes receive no rule-shift test entries because all their rules were retained in training.

\mypar{Rule-Shift.}
A conventional random train/test split cannot tell us whether the model learned the byte/instruction patterns related to a functionality or simply memorized the byte patterns that trigger a specific CAPA rule.
The rule-shift set addresses this: every one of its test functions is labeled by a rule the model never saw during training, so accuracy on this set measures how well the model transfers from the rules it did see to the ones it did not.

\mypar{Baseline Models.}
To evaluate \sys{}'s performance for functionality classification, we first compare it against \sys{}-N.
We then compare \sys{} against two additional baselines: Malconv2~\cite{raff2021classifying} and a multi-layer Perceptron (MLP) model, which serves as an ablation study to assess the usefulness of the sequential transformer architecture.
For the MLP model, the weights of the byte-level token embeddings are randomly initialized and updated during training.
The tokenized sequence is then transformed into a fixed-size function embedding by applying average pooling, which is subsequently fed into an MLP model for classification.
Both Malconv2 and the MLP are trained directly on raw function bytes with a maximum input length of 1,048 bytes, whereas \sys{} uses the first 1,024 tokens.
The detailed training setup for \sys{} and all the baselines are explained in Appendix~\ref{app:finetuning_setup}

\mypar{Results.}
\begin{table*}[t]
\centering
\caption{Top-$k$ accuracy for functionality classification under the normal and rule-shift test splits. We report Top-1, Top-3, and Top-5 accuracy for the baseline models and \ssys{} 1,024.}
\label{tab:functionality_classification_baseline}
\resizebox{\textwidth}{!}{%
\begin{tabular}{|l|ccc|ccc|}
\toprule
\multirow{2}{*}{\textbf{Model}} & \multicolumn{3}{c|}{\textbf{Normal}} & \multicolumn{3}{c|}{\textbf{Rule Shift}} \\
\cmidrule{2-7}
& {\textbf{Top-1} (\%)} & {\textbf{Top-3} (\%)} & {\textbf{Top-5} (\%)} & {\textbf{Top-1} (\%)} & {\textbf{Top-3} (\%)} & {\textbf{Top-5} (\%)} \\
\midrule
\rowcolor{gray!20}
\ssys{}-N 1,024 & 83.70 & 89.18 & 91.15 & 18.85 & 24.46 & 31.27 \\
\midrule
Malconv2 & 67.52 & 74.53 & 77.83 & 4.00 & 8.52 & 12.34 \\
\midrule

\rowcolor{gray!20}
MLP & 76.63 & 85.50 & 88.75 & 13.06 & 22.55 & 28.51 \\
\midrule

\ssys{} 1,024 & \textbf{91.80} & \textbf{95.27} & \textbf{96.28} & \textbf{29.10} & \textbf{40.21} & \textbf{44.55} \\
\bottomrule
\end{tabular}%
}
\end{table*}

Table~\ref{tab:functionality_classification_baseline} reports Top-1, Top-3, and Top-5 accuracy under the normal and rule-shift test splits.
\sys{}-1,024 achieves Top-1/Top-3/Top-5 of 91.80/95.27/96.28\% on the normal split and 29.10/40.21/44.55\% on the rule-shift split, outperforming all baselines in both settings.
Because every test function in the rule-shift split is labeled by a CAPA rule the model never saw during training, its rule-shift accuracy measures transfer from seen rules to unseen ones, and the gap from the normal split quantifies how much of the within-distribution accuracy was tied to rule-specific byte patterns rather than functionality-level structure.
Comparing \sys{}-1,024 with its non-pretrained counterpart \sys{}-N 1,024 isolates the effect of pretraining: it adds 8.10 Top-1 points on the normal split (91.80 vs.\ 83.70) and 10.25 points on the rule-shift split (29.10 vs.\ 18.85), and the larger rule-shift gain indicates that pretraining contributes more to transfer across unseen labeling rules than to within-distribution accuracy alone.
Among the remaining baselines, the MLP consistently outperforms Malconv2, indicating that architectures optimized for whole-binary malware classification do not automatically transfer to localized function-level behavior labeling.

\mypar{Per-Functionality Behavior.}
The per-class Top-1 accuracy of \sys{}-1,024 on both splits is shaped by what a CAPA rule matches on and how much of that information is visible in raw function bytes.
CAPA rules fall broadly into two categories: rules that match on a specific symbolic API name (e.g., \texttt{CreateFileA}, \texttt{WSAStartup}) and rules that match on algorithm-specific instruction patterns (e.g., the computational core of a hash).
Our model consumes raw function bytes without resolved imports, so algorithm-specific patterns are directly observable, while symbolic API names are not --- the model can only learn a surrogate from the surrounding argument setup and call-site patterns.

\noindent\textbf{Normal split.}
Algorithm-anchored functionalities reach near-perfect per-class accuracy: \emph{get disk information via IOCTL}, \emph{decompress data using aPLib}, \emph{hash data with MD5}, and \emph{hash data using fnv} sit at the top because their distinguishing computation is directly visible in the bytes.
API-anchored functionalities still score highly when their surrogate is stable. For example, widely-used primitives with consistent calling conventions, such as \emph{terminate process} and \emph{initialize Winsock library}, because the argument setup and call pattern around the API acts as a reliable proxy for the rule's API match.
Accuracy is lowest for categories that are either too rare or too broad. Rare categories (under 100 examples) don't give the model enough data to learn from, while broad categories (like HTTP User-Agent string) have too many variations to form a single, clear pattern.

\noindent\textbf{Rule-shift split.}
The two kinds of rules behave differently under rule-shift.
Algorithm-anchored functionalities transfer cleanly.
\emph{hash data using fnv} and \emph{compute adler32 checksum} retain per-class accuracy close to their normal-split values, because the computational core the model relies on is present in the bytes.
Transfer is bound for API-anchored functionalities such as file I/O, registry access, and process management.
The argument setup and call-site pattern the model learned for these classes is tied to the specific API the original rule matched on.
When a different rule labels the same functionality through a different API, that pattern no longer applies, and the prediction no longer points to the correct class.
This gives a representational reading of the gap: pretrained code representations carry algorithmic computation across unseen rules, while API-anchored functionalities transfer only as far as the raw-byte context around their API calls allows.

\begin{takeawaybox}
\sys{}-1,024 reaches 91.80\% Top-1 on the normal split and 29.10\% Top-1, reaching 44.55\% Top-5 accuracy on the rule-shift split, showing that its representations transfer to functionalities labeled by CAPA rules the model never saw during training. 
The transfer is driven by algorithmic structure that raw function bytes directly expose, which pretrained code representations are able to capture across labeling rules.
\end{takeawaybox}

\subsection{Malware Detection}
\label{sec:malware_detection}
In this section, we compare the performance of \sys{} with four baseline models for malware detection.
Malconv2~\cite{raff2021classifying} is a gated convolutional neural network that processes the entire executable as a raw byte sequence, using temporal max-pooling to handle variable-length inputs with constant memory.
MalGraph~\cite{ling2022malgraph} is a hierarchical graph neural network that constructs control flow graphs (CFGs) for each function and a function call graph (FCG) across functions, applying GraphSAGE layers to learn representations at both the intra-function and inter-function levels before predicting maliciousness.
EmberGBM and EmberNN both operate on 2,381 handcrafted Ember features~\cite{anderson2018ember} extracted from PE headers and sections; EmberGBM uses a gradient-boosted decision-tree model, while EmberNN uses a multi-layer perceptron.
The detailed training setup for \sys{} and all the baselines are provided in Appendix~\ref{app:finetuning_setup}.
\begin{table*}[t]
\centering
\caption{Malware detection under temporal drift. TPR@0.1\% is the true-positive rate at FPR~=~0.001.
For brevity, we refer to \sys{} as \ssys{}.
\textbf{Bold} = best per column; \underline{underline} = second best.}
\label{tab:concept_drift_packed}
\resizebox{\textwidth}{!}{%
\begin{tabular}{|c|c|c!{\vrule width 1.1pt}c|c!{\vrule width 1.1pt}c|c!{\vrule width 1.1pt}c|c!{\vrule width 1.1pt}c|c!{\vrule width 1.1pt}c|c|}
\toprule
\textbf{Model}
& \multicolumn{6}{c!{\vrule width 1.1pt}}{\textbf{Train: D1}}
& \multicolumn{4}{c!{\vrule width 1.1pt}}{\textbf{Train: D1+D2}}
& \multicolumn{2}{c|}{\textbf{Train: D1+D2+D3}} \\
\cmidrule{2-13}
\textbf{Test Set}
& \multicolumn{2}{c!{\vrule width 1.1pt}}{\textbf{D1}}
& \multicolumn{2}{c!{\vrule width 1.1pt}}{\textbf{D2}}
& \multicolumn{2}{c!{\vrule width 1.1pt}}{\textbf{D3}}
& \multicolumn{2}{c!{\vrule width 1.1pt}}{\textbf{D2}}
& \multicolumn{2}{c!{\vrule width 1.1pt}}{\textbf{D3}}
& \multicolumn{2}{c|}{\textbf{D3}} \\
\cmidrule{2-13}
& \textbf{TPR@0.1\%} & \textbf{AUC}
& \textbf{TPR@0.1\%} & \textbf{AUC}
& \textbf{TPR@0.1\%} & \textbf{AUC}
& \textbf{TPR@0.1\%} & \textbf{AUC}
& \textbf{TPR@0.1\%} & \textbf{AUC}
& \textbf{TPR@0.1\%} & \textbf{AUC} \\
\midrule

\rowcolor{gray!20}
EmberGBM
& \textbf{86.40} & \textbf{0.999} & \textbf{92.10} & \textbf{0.999} & \textbf{44.80} & \textbf{0.995}
& \textbf{96.70} & \textbf{1.000} & \textbf{44.50} & \textbf{0.993}
& \textbf{53.50} & \textbf{0.992} \\

Ember NN
& 25.00 & \underline{0.993} & 13.90 & \underline{0.996} & 21.90 & \underline{0.983}
& 21.50 & \underline{0.996} & 39.00 & 0.985
& 37.68 & 0.927 \\
\midrule

\rowcolor{gray!20}
MalGraph
& 15.70 & 0.986 & 57.60 & 0.995 & 14.70 & 0.974
& 52.20 & \underline{0.996} & 19.70 & 0.982
& 13.40 & 0.950 \\

\midrule

Malconv2
& 12.30 & 0.963 & 6.60 & 0.977 & 3.60 & 0.950
& 3.10 & 0.979 & 2.20 & 0.963
& 0.09 & 0.900 \\

\midrule

\rowcolor{gray!20}
\ssys{} 1024
& \underline{39.23} & 0.991 & \underline{72.28} & 0.994 & \underline{29.30} & \underline{0.983}
& \underline{82.76} & 0.991 & \underline{40.18} & \underline{0.988}
& \underline{40.80} & \underline{0.96} \\
\bottomrule
\end{tabular}%
}
\end{table*}

\mypar{Concept Drift Evaluation.}
We evaluate temporal drift by incrementally expanding the training set with newer malware cohorts, using the temporal splits and dataset defined in Section~\ref{sec:eval_setup}.
In the first setting, models train on $D_1$ and are tested on $D_1$, $D_2$, and $D_3$.
In the second, $D_2$ is added to training and models are tested on $D_2$ and $D_3$.
In the third, all three cohorts are used for training and models are tested on $D_3$.
In each setting, the benign train, validation, and test splits remain fixed, and each test set pairs the malicious cohort's test split with the benign test split.
Benign samples are not sorted chronologically, as prior work has shown that performance decay under temporal drift is primarily driven by evolution in the malicious class, while benign distributions remain comparatively stable~\cite{barbero2022transcending}.
This simulates a deployment scenario where the detector is periodically retrained on fresher malware.

\mypar{Metrics.}
We report AUC and TPR at FPR~=~0.001 (0.1\%).
AUC summarizes the classifier's ability to rank malicious samples above benign ones across all thresholds, providing a threshold-agnostic view of class separation under our balanced evaluation.
TPR at FPR~=~0.001 measures the fraction of malicious samples detected at a strict low-FPR operating point.
In deployment, benign files vastly outnumber malicious ones, so even small false-positive rates produce large volumes of false alerts; recent evaluation guidance therefore recommends reporting TPR at low FPR to reflect realistic operational performance~\cite{arp2022and}.

\mypar{Detection Results.}
Table~\ref{tab:concept_drift_packed} summarizes the performance across all temporal concept-drift settings.
At this strict operating point, EmberGBM is the strongest model overall, achieving the highest TPR@0.1\% in every setting.
\sys{}-1024 achieves the second-highest TPR@0.1\% across all settings, outperforming every neural baseline: it reaches 39.23\% on D1$\rightarrow$D1, 72.28\% on D1$\rightarrow$D2, and 40.80\% on D1{+}D2{+}D3$\rightarrow$D3.
Among the remaining neural baselines, MalGraph is strongest on D1$\rightarrow$D2 and D1{+}D2$\rightarrow$D2, EmberNN leads on the remaining four splits, and Malconv2 degrades most sharply under this low-FPR constraint.
This highlights how difficult malware detection becomes when the false-positive budget is constrained to FPR~=~0.001 (0.1\%), especially under temporal shift.

\mypar{Qualitative Error Analysis.}
\sys{} and EmberGBM differ fundamentally in their input representations --- code-section bytes versus structural PE features --- which suggests the two models may capture complementary signals.
To investigate, we manually inspected small sets of disagreements between \sys{}-1024 and EmberGBM;
Appendix~\ref{app:malware_detection_error_analysis} provides the detailed analysis.
When \sys{} correctly classified samples missed by EmberGBM, EmberGBM's errors were often associated with metadata features such as file size, digital signatures, certificate tables, debug directories, and section-permission layouts.
These metadata fields are decoupled from runtime behavior and can be added post-compilation without altering execution, whereas \sys{} reads the code segment directly and stays anchored to what the binary does.
Conversely, samples missed by \sys{} but correctly classified by EmberGBM often included packed, polymorphic, adware, or dropper-like files where non-code structural features remained informative.
This supports a narrower conclusion than model superiority: code-section features and PE-structure features are complementary, and an ensemble may be better suited for strict low-FPR operation than either feature family alone.

\mypar{Discussion.}
These results highlight an important distinction between malware \emph{detection} and malware \emph{analysis}.
For strict low-FPR file-level detection, structural PE features remain highly effective, although the error analysis above suggests that part of this advantage comes from metadata fields and layout artifacts that may be easier to manipulate post-compilation without altering execution.
Among the neural baselines we evaluate, \sys{} consistently achieves the highest TPR@0.1\% and, unlike feature-engineering approaches, learns a reusable code-centric representation that also transfers to API-call prediction and function-level behavior labeling, tasks that require finer-grained prediction rather than only file-level discrimination.

\begin{takeawaybox}
At FPR~=~0.001, malware detection remains challenging even for models with strong overall ranking performance. The results and qualitative error analysis suggest that code-section features and PE-structure features capture complementary signals, rather than one consistently dominating the other.
\end{takeawaybox}

\subsection{Comparison with other Foundation Models}

\label{sec:eval_comparison_sota}

To position \sys{} within the broader landscape of binary foundation models, we compare its performance against XDA~\cite{pei2020xda} and large malware language models (LMLMs) from recent work~\cite{kurlandski2026beyond}. 
This comparison highlights three key design choices: pretraining data domain (benign vs. malware), tokenization granularity (byte-level vs. BPE), and sequence modeling approach (whole-file long sequences versus fixed-length chunking with aggregation).

\mypar{Baselines.}
XDA is pretrained on compiler-optimized binaries from the SPEC CPU benchmarks~\cite{speccpu2006,speccpu2017} using byte-level (256-token) encoding. Since these benchmarks differ substantially from real-world malware distributions, XDA serves as a baseline for non-malware-aligned pretraining.
The LMLM baselines use BPE tokenization on executable sections with a 16,384 vocabulary size, much larger than similar to \sys{}, but employ long-sequence architectures (Bidirectional Mamba and HRRFormer) rather than chunked aggregation~\cite{kurlandski2026beyond} for malware detection. 
We fine-tune both XDA and the LMLM baselines using identical data splits and hyperparameters as \sys{} across API-call prediction, functionality classification, and malware detection. Results are summarized in Table~\ref{tab:comparison_language_models}.
 
\begin{table}[t]
\centering
\caption{Comparison with XDA and EXE-based LMLM baselines~\cite{kurlandski2026beyond}. TPR@0.1\% is the true-positive rate at FPR~=~0.001. Dashes indicate unavailable values.}
\label{tab:comparison_language_models}
\resizebox{\textwidth}{!}{%
\begin{tabular}{|l|l|cc|cc|cc|c|}
\toprule
\textbf{Model} & \textbf{Pretraining Data} & \multicolumn{2}{c|}{\textbf{API Call Prediction}} & \multicolumn{2}{c|}{\textbf{Func. Class. Normal}} & \multicolumn{2}{c|}{\textbf{Func. Class. RuleShift}} & \textbf{Malware Detect.} \\
\cmidrule{3-9}
& & \textbf{Top-3} & \textbf{Top-5} & \textbf{Top-1} & \textbf{Top-3} & \textbf{Top-1} & \textbf{Top-3} & \textbf{TPR@0.1\%} \\
\midrule

XDA & ELF text segments & 82.52 & 91.27 & 88.34 & 92.60 & 22.81 & 33.00 & 30.17 \\

\rowcolor{gray!20}
LMLM Bi-Mamba & PE executable segments & 77.76 &  83.98  & 77.00 & 85.97 & 12.11 & 21.46 & 35.08 \\

LMLM Bi-HRRFormer & PE executable segments & 44.32 & 49.90 & 34.35 & 62.10 & 9.30 & 17.00 & 0.40 \\

\rowcolor{gray!20}
MALPT 1024 & PE executable segments & \textbf{92.17} & \textbf{93.01} & \textbf{91.80} & \textbf{95.27} & \textbf{29.10} & \textbf{40.21} & \textbf{39.23} \\
\bottomrule
\end{tabular}%
}
\end{table}

Table~\ref{tab:comparison_language_models} summarizes the comparison results.
\sys{}-1024 achieves the strongest results across all the downstream tasks.  

\mypar{Comparison with XDA.}
Relative to XDA, \sys{}-1024 improves API-call prediction from 82.52\% to 92.17\% in Top-3 accuracy and from 91.27\% to 93.01\% in Top-5 accuracy.
It also improves functionality classification under both the normal split (Top-1/Top-3: 88.34/92.60 to 91.80/95.27) and the rule-shift split (22.81/33.00 to 29.10/40.21).
Because XDA uses benchmark-pretrained byte-level representations rather than malware-aligned code pretraining, this comparison supports the importance of matching both pretraining data and tokenization to the malware domain.

\mypar{Comparison with Long-Sequence LMLMs.}
The EXE-based Bi-Mamba and Bi-HRRFormer baselines employ long-sequence architectures for file-level malware detection.
\sys{}-1024 reaches 39.23\% TPR@0.1\%, above the 35.08\% of Bi-Mamba and substantially above the 0.40\% of Bi-HRRFormer, showing that a fixed-length encoder combined with chunk-level aggregation can outperform long-sequence modeling at a strict low-FPR operating point.
The gap widens at finer granularities: on functionality classification, \sys{}-1024 substantially exceeds both LMLM baselines on the normal and rule-shift splits, suggesting that architectures optimized for whole-file classification do not automatically provide strong inductive bias for function-level behavior labeling.
Finally, compared with XDA's byte-level vocabulary and the 16,384-token BPE used by the LMLMs, \sys{} performs best overall at 1,024 tokens, supporting the selection of tokenization granularity based on downstream performance rather than defaulting to the largest vocabulary (see Section~\ref{sec:eval_vocab_size}).

\begin{takeawaybox}

Compared with XDA and raw-byte-based Bi-Mamba/Bi-HRRFormer LMLMs, \sys{} achieves the strongest performance across all downstream tasks. 
Together, these comparisons show that malware-aligned pretraining, downstream-validated tokenization, and chunk-level document aggregation are strong alternatives to purely whole-segment long-sequence modeling.
\end{takeawaybox}

\subsection{Effect of Vocabulary Size}
\label{sec:eval_vocab_size}

A core hypothesis of our work is that using a BPE tokenization scheme yields better downstream performance than the byte-level encoding used by previous works like XDA for malware analysis tasks.
To validate this hypothesis, we evaluate \sys{} across five different vocabulary sizes: 256, 512, 1,024, 2,048, and 4,096, generated via Byte-Pair Encoding (BPE). We consider the byte-level \sys{}-256 model as our baseline.
To ensure a controlled comparison, we pretrained all model variants on the same dataset of \totalPretrainingSamples{} malicious binaries for an identical duration of 150,000 training steps.
Furthermore, during the evaluation phase, we fine-tuned each model using consistent dataset splits and hyperparameters across every downstream task to isolate the specific impact of vocabulary size.
Table~\ref{tab:ablation_vocab} summarizes the performance across all three downstream tasks.
\begin{table*}[t]
\centering
\caption{Effect of BPE vocabulary size on downstream performance. TPR@0.1\% is the true-positive rate at FPR~=~0.001.}
\label{tab:ablation_vocab}
\resizebox{\textwidth}{!}{%
\begin{tabular}{|l|cc|cc|cc|c|}
\toprule
\multirow{2}{*}{\textbf{Model Type}} 
& \multicolumn{2}{c|}{\textbf{API call Prediction}} 
& \multicolumn{2}{c|}{\textbf{Func. Class. Normal}} 
& \multicolumn{2}{c|}{\textbf{Func. Class. RuleShift}} 
& \textbf{Malware Detection} \\
\cmidrule{2-8}
& {\textbf{Top-1} (\%)} & {\textbf{Top-5} (\%)}
& {\textbf{Top-1} (\%)} & {\textbf{Top-3} (\%)}
& {\textbf{Top-1} (\%)} & {\textbf{Top-3} (\%)}
& {\textbf{TPR@0.1\%}} \\
\midrule

\rowcolor{gray!20}
MalwarePT 256  & 85.91 & 90.76 & 89.53 & 93.39 & 26.98 & 34.32 & 38.24 \\

MalwarePT 512  & \textbf{88.04} & 93.00 & 90.43 & 94.35 & 25.41 & 34.62 & 38.61 \\

\rowcolor{gray!20}
MalwarePT 1024 & 87.83 & \textbf{93.01} & \textbf{91.80} & \textbf{95.27} & \textbf{29.10} & \textbf{40.21} & 39.23 \\

MalwarePT 2048 & 86.62 & 92.48 & 90.29 & 94.32 & 26.40 & 36.36 & \textbf{40.75} \\

\rowcolor{gray!20}
MalwarePT 4096 & 86.51 & 91.73 & 89.85 & 94.11 & 25.65 & 35.06 & 37.84 \\
\bottomrule
\end{tabular}%
}
\end{table*}

\mypar{Token Granularity Improves Performance.}
Across all downstream tasks, we observe a performance improvement as we increase the vocabulary size from the baseline of 256 tokens to 512 and 1,024 tokens.
For instance, Top-1 API accuracy rises from 85.91\% at 256 tokens to 87.83\% at 1,024; functionality Top-3 improves on both the normal split (93.39\% to 95.27\%) and the rule-shift split (34.32\% to 40.21\%); and malware-detection TPR@0.1\% rises from 38.24\% to 39.23\% (Table~\ref{tab:ablation_vocab}).
This confirms that merging raw bytes into larger tokens allows the model to use local code patterns (such as common instructions, API-call stubs, or function prologues) more effectively than treating every byte as an independent unit.
In addition, BPE reduces the sequence length required to represent the same byte sequence, enabling the model to attend to longer-range dependencies within its fixed context window.

\mypar{Diminishing Returns at Larger Vocabularies.}
Interestingly, performance does not scale linearly with increase in vocabulary size; beyond 1,024 tokens, we observe diminishing returns or even a slight degradation.
For API-call prediction, Top-1 accuracy drops to 86.62\% at 2,048 tokens and 86.51\% at 4,096 tokens, while Top-5 accuracy drops from 93.01\% at 1,024 tokens to 92.48\% and 91.73\%, respectively.
A similar pattern can be observed for functionality classification, where the normal-split Top-3 accuracy drops from 95.27\% at 1,024 tokens to 94.32\% at 2,048 and 94.11\% at 4,096, while the rule-shift Top-3 accuracy falls from 40.21\% to 36.36\% and 35.06\%, respectively.
For malware detection, TPR@0.1\% peaks at 2,048 tokens (40.75\%) before dropping to 37.84\% at 4,096, showing that diminishing returns set in at different vocabulary sizes depending on the task.
We attribute this behavior to the increased sparsity of the token distribution at larger vocabulary sizes.
As the vocabulary grows to 4,096, the tokenizer creates specialized tokens for rare byte sequences.
These rare tokens appear infrequently during pretraining, limiting embedding quality and ultimately degrading fine-tuning performance.
Based on these results, we identify 1,024 tokens as the best overall vocabulary, with 2,048 offering only a marginal TPR@0.1\% advantage that does not carry over to other tasks.

\begin{takeawaybox}
Larger BPE vocabularies improve token granularity up to a point, but gains do not scale monotonically. The results support 1,024 tokens as the strongest overall tradeoff, with 512 remaining competitive on selected tasks.
\end{takeawaybox}

\section{Security Discussion}
\label{sec:discussion}
We discuss more possible ways of evading our malware analysis framework. 

\mypar{Code Segment Evasion.}
In this work, code segment is defined as the segment in an executable file consisting program's entry point (offset of the first executed instruction). 
However, malware authors can undermine this assumption by creating a dummy code segment, redirecting the entry point, and using jump instructions to point execution to the actual code segment.
Currently, our code segment identification algorithm cannot handle such scenarios and will fail to extract the raw bytes of the actual code segment. 
This limitation arises not from the ML model but from the pre-processing tool used to identify the code segment. 
A potential solution to this limitation is to include all segments marked as readable and executable, along with the entry segment, as part of the raw code segment and extract the raw bytes from all of them, or conduct a manual verification of the extracted code segment. 

\mypar{Packed Code.}
Given the static nature of our approach, adversaries can develop novel and adaptive packing techniques that render our analysis ineffective. 
One potential solution to address this challenge is similar to recent work ~\cite{vasan2024deepcapa}, by leveraging dynamic artifacts, such as memory snapshots, instead of relying only on the executable file to extract code segments. 
These memory snapshots captured after the packed executable has unpacked itself can be used to obtain the original code segment, allowing for a more effective detection.

\mypar{Context Length.}
We pretrain \sys{} on tokenized raw bytes from the code segment, using a context length of \seqLength{} tokens (\contentLength{} content tokens plus delimiters).
This context length, combined with Rotary Position Embeddings (RoPE), allows the model to capture patterns at the instruction, function, and segment levels, as demonstrated by the strong performance of \sys{}.
Moreover, RoPE enables context length extension at fine-tuning time without re-interpolation, should longer sequences be needed for specific tasks.
In addition, our BPE tokenizers (512--4,096 vocabulary) enable each token to represent more information and thus capture longer code sequences within the same token budget.
For example, our 4,096-vocabulary BPE tokenizer can encode up to approximately 1,500 bytes in a \seqLength{}-token sequence, offering substantially longer contextual coverage than byte-level tokenization.
However, we acknowledge that certain malware behaviors may span beyond the \seqLength{}-token limit, potentially impacting \sys{}’s ability to capture long-range dependencies.

\mypar{Computational Cost.}
As a transformer-based foundation model, \sys{} incurs higher per-file latency than lightweight static analyzers.
Preprocessing (code-section extraction and BPE tokenization) takes approximately 14~ms per file, and encoder inference on a single L40S GPU for malware detection averages 3.3~s per file (median 783~ms), dominated by the number of 1,024-token chunks per binary (median 33, mean 152).
By comparison, the Ember-based models complete feature extraction and prediction in approximately 50~ms per file, and Malconv2 averages 54~ms.
MalGraph, however, requires IDA Pro disassembly for CFG and FCG construction, averaging 4.1 minutes per file — substantially slower than \sys{}.
We view \sys{} as suited for backend malware-analysis pipelines, where deeper code understanding justifies higher per-file cost rather than endpoint-level filtering.
Appendix~\ref{app:computational_cost} provides detailed cost breakdowns for preprocessing, inference across all baselines, fine-tuning, and pretraining.
Appendix~\ref{app:model_architecture} lists the full model architecture and pretraining hyperparameters.

\section{Related Work}
\label{sec:related_work}
\mypar{Malware Classification.} 
Malware classification for PE files typically involves binary classification to detect malicious and benign samples or performing multi-class classification for malware family classification.
Existing methods fall into two categories: feature-engineering-based and raw-byte-based approaches.

Feature-engineering-based methods~\cite{anderson2018ember,shafiq2009pe,santos2013opcode,saxe2015deep,rathore2018malware,ahmadi2016novel} extract static features such as file headers, imports, strings, and bytecode distributions, which are then processed by traditional ML-based models (e.g., Naive Bayes, decision trees, SVMs) or deep learning models (e.g., MLPs) to perform detection.
However, their reliance on task-specific handcrafted features limits applicability across different downstream tasks.

Raw-byte-based approaches eliminate this dependence by processing entire executables as a whole ~\cite{raff2018malware,raff2021classifying,vasan2020imcfn,kalash2018malware,kumar2018malicious,bhodia2019transfer,makandar2017malware,yue2017imbalanced}.
Although these models perform well, they are typically optimized for whole-executable classification.
They are not designed for transferable analysis across multiple malware-analysis granularities.
Feature-engineering pipelines and raw-byte CNNs also remain tied to task-specific objectives, which limits reuse across downstream tasks.

\mypar{Foundation Models for Binary Code Analysis.}
Recent work has explored foundation models for binary code analysis, particularly in the contexts of binary similarity detection~\cite{koo2021semanticawarebinarycoderepresentation,ahn2022practical}, reverse engineering~\cite{pei2020xda}, and code region identification~\cite{benkraouda2024you}. DeepSemantic~\cite{koo2021semanticawarebinarycoderepresentation} and BinShot~\cite{ahn2022practical} train BERT-based models on assembly code for binary similarity tasks, which differ in both objective and input modality from our work.
Both XDA~\cite{pei2020xda} and ~\cite{benkraouda2024you} operate on raw bytes and are optimized for instruction-level tasks such as function boundary detection and for distinguishing code from non-code regions.
However, \sys{} also performs analysis at the function and segment levels by aggregation of multiple sequence representations and the design of specialized classification heads.

\section{Conclusions}
\label{sec:conclusion}
In this paper, we introduced \sys{}, a binary-level foundation model for malware analysis pretrained on malicious Windows PE code-section bytes, and evaluated it on API call prediction, function-level behavior-label prediction, and malware detection. The results show that malware-specific pretraining transfers strongly to token- and function-level tasks, while byte-pair tokenization improves the effective context budget beyond the byte-level baseline and yields the strongest overall tradeoff at 1,024 tokens.



Taken together, these findings position pretraining domain and tokenization granularity as first-order design choices for malware foundation models.

\subsubsection*{Acknowledgements.}
This material is based upon work supported by the National Science Foundation under grant no. 2229876 and is supported in part by funds provided by the National Science Foundation, by the Department of Homeland Security, and by IBM.

Any opinions, findings, and conclusions or recommendations expressed in this material are those of the author(s) and do not necessarily reflect the views of the National Science Foundation or its federal agency and industry partners.

\clearpage
\bibliographystyle{splncs04}
\bibliography{refs}

\clearpage
\section{Statement on Data Availability}
We are committed to maximizing the reproducibility of our work, and upon publication we will release all source code and pretrained model weights for every version of \sys—including the 256–4096 vocabulary models as well as the benign-only and mixed-domain variants—via an anonymized GitHub repository (link to be provided in the camera-ready version).

We will also publish the full list of malware and benign sample hashes used in our dataset, enabling independent retrieval of these files from trusted services such as VirusTotal. 

\clearpage
\appendix
\renewcommand{\theHsection}{appendix.\Alph{section}}

\section{Finetuning Setup}
\label{app:finetuning_setup}
We fine-tune \sys{} for three downstream tasks: API call prediction, functionality classification, and malware detection.
In this section, we first describe the end-to-end fine-tuning setup used across all tasks, then detail the task-specific hyperparameters for \sys{} and all baselines.

\mypar{End-to-End Fine-Tuning.}
All downstream \sys{} models are optimized end-to-end: the pretrained encoder and the task-specific output head are updated jointly during downstream training rather than through parameter-efficient adapters. We use the same mixed-precision training regime described in Section~\ref{sec:eval_setup} and select the checkpoint with the lowest validation loss for final test evaluation. This keeps the adaptation protocol consistent with the end-to-end optimization used for the neural baselines.

\mypar{API call Prediction.}
To enable direct token-level prediction of API calls, we extend \sys{}'s pretrained vocabulary of size $U \in \{512, 1024, 2048, 4096\}$ by appending $A = 1280$ API call tokens, resulting in an expanded vocabulary of size $U + A$.
This corresponds to adding $A$ new rows to the embedding layer's matrix.
A new model is initialized with this extended embedding matrix: the first $U$ rows are copied from the pretrained model, while the remaining $A$ rows are randomly initialized and updated during fine-tuning.
We also replace the original classification head with a new one consisting of $A$ output units.
During end-to-end fine-tuning, the expanded embedding matrix, the output head, and the pretrained backbone are optimized jointly.
All models are fine-tuned for 500 steps using a batch size of 256.
For \sys{}-N we use the same model architecture as \sys{}, and do not load weights from the pretrained stage.
\sys{} is fine-tuned for 5 epochs, while \sys{}-N is trained for 20 epochs.

\mypar{Functionality Classification.}
We train \sys{} and \sys{}-N using the same underlying architecture and compare against two baseline models: Malconv2 and a multi-layer perceptron (MLP).

For \sys{}, we initialize with pretrained weights and replace the pretraining classification head with a new one consisting of \(F\) classification heads.
These models are fine-tuned for 5 epochs.
\sys{}-N is initialized from scratch and trained for 20 epochs.

Malconv2 is configured as described in the original work: batch size of 128, hidden size of 8, convolutional filter size of 256, and stride of 64.
The MLP consists of an embedding layer, average pooling, and three feed-forward layers.
We use an embedding dimension of 768; the first two feed-forward layers have size 3072, and the final layer outputs \(F\) classification scores.
We apply ReLU activation between layers and use a dropout of 0.2 between the first and second layers.
For functionality classification, both Malconv2 and the MLP are trained directly on raw function bytes with a maximum input length of 1,024 bytes, whereas \sys{} uses the first 1,024 tokens.
Both Malconv2 and the MLP model are trained for 20 epochs.

\mypar{Malware Detection}
We compare the performance of \sys{} and three baseline models: EmberGBM, EmberNN, and Malconv2.
For \sys{}, we replace the pretraining heads with a malware detection head composed of a multi-head self-attention layer followed by a feed-forward classification module.
The attention layer uses 12 heads, and the feed-forward component consists of two layers with hidden dimensions of 3072 and 2, respectively.
We fine-tune both models for 10 epochs.

For EmberGBM, we use a gradient-boosted decision tree model implemented with the LightGBM framework.
The model is configured with 1000 boosting iterations, a learning rate of 0.05, and uses the binary classification objective.
To increase model capacity, we set the number of leaves to 2048 and the maximum tree depth to 15.
Additionally, we apply a minimum data-in-leaf threshold of 50 and set the feature fraction to 0.5 to encourage feature subsampling during training.

EmberNN is implemented as a deep feed-forward neural network that takes as input the same feature representation used by EmberGBM.
The architecture consists of four fully connected layers: the first projects the input to a 4000-dimensional space, followed by ReLU activation, batch normalization, and a dropout rate of 0.5.
The second layer reduces the representation to 2000 dimensions and applies the same activation, normalization, and dropout sequence.
This is followed by a third layer projecting to 100 dimensions, again followed by ReLU, batch normalization, and dropout.
Finally, the output layer maps to the number of target classes. This network is trained using cross-entropy loss.

Malconv2 uses the same configuration as described in the functionality classification task: a convolutional architecture with a hidden layer size of 8, a convolutional filter width of 256, and a stride of 64, with a batch size of 128.
Both Malconv2 and EmberNN are trained for 30 epochs.

For all downstream experiments involving neural network models, we select the model checkpoint with the lowest validation loss for final evaluation on the test set.

\section{Ablation Study: Classification Head Architectures}
\label{app:ablation_study}

We conduct an ablation study comparing three end-to-end fine-tuning approaches for the malware detection task, differing only in their final classification head architecture. All models use the same pretrained encoder and are trained end-to-end with the encoder weights updated jointly with the classification head.

\mypar{Classification Head Variants.}
We evaluate three classification head designs:

\begin{enumerate}
\item \textbf{CNN-1D Head:} A 1D convolutional layer followed by max-pooling, which learns local spatial patterns in the aggregated sequence representation. Configuration: kernel size 8, stride 1, max-pooling kernel size 2, max-pooling stride 2.

\item \textbf{Average Pooling Head:} A simple aggregation baseline that averages sequence representations across the sequence dimension, followed by a two-layer feed-forward network (3072 hidden units, then 2 output units). This provides a minimal, non-parametric aggregation baseline.

\item \textbf{Multihead Attention + RoPE Head:} A learned attention-based aggregator using 12 attention heads with rotary position embeddings (RoPE), followed by a feed-forward module (3072 hidden units, then 2 output units). This approach learns data-dependent aggregation weights rather than applying fixed pooling operations.
\end{enumerate}

\mypar{Evaluation Setup.}
All three variants are trained using the same training, validation, and test sets from Section~\ref{sec:malware_detection} (malware samples D1 paired with benign samples, 50/10/40 split, evaluated on the combined test set). Models are fine-tuned for 10 epochs with batch size 64, and we select the checkpoint with lowest validation loss for evaluation.

Table~\ref{tab:ablation_head_architectures} summarizes the performance of the three classification-head variants.
\begin{table}[t]
\centering
\caption{Ablation Study: Comparison of classification head architectures for malware detection using end-to-end fine-tuning on \sys{} with 1024-token vocabulary. Reported metrics are AUC and TPR@0.1\% (true-positive rate at FPR = 0.001).}
\label{tab:ablation_head_architectures}
\begin{tabular}{|l|S[table-format=2.2]|S[table-format=2.2]|}
\toprule
\textbf{Classification Head} & \textbf{AUC (\%)} & \textbf{TPR@0.1\% (\%)} \\
\midrule
Average Pooling & 95.20 & 16.28 \\
\rowcolor{gray!20}
CNN-1D & 98.70 & 35.41 \\
Multihead Attention + RoPE & \textbf{99.10} & \textbf{39.23} \\
\bottomrule
\end{tabular}
\end{table}

\mypar{Results and Analysis.}
The multihead attention-based head with RoPE achieves the strongest performance (AUC = 99.10, TPR@0.1\% = 39.23), outperforming the CNN-1D head (AUC = 98.70, TPR@0.1\% = 35.41) by 3.8 percentage points in TPR at the strict FPR constraint. The average pooling baseline (AUC = 95.20, TPR@0.1\% = 16.28) performs substantially worse, indicating that learned aggregation mechanisms provide meaningful improvements over fixed pooling for this task.

These results suggest that the attention mechanism's ability to learn sample-specific importance weights across the sequence dimension is beneficial for malware detection, particularly when operating under strict false-positive rate constraints where confidence calibration matters.


\section{Ablation for Pretraining Dataset}
\label{sec:eval_ablation_dataset}
\begin{table*}[t]
\centering
\caption{Impact of pretraining data domain on downstream performance for \sys{}-4096.
We compare models pretrained exclusively on malicious code, benign code, and a mixed dataset.
We report Top-1 and Top-5 accuracy for API call prediction,
Top-1 and Top-3 accuracy for functionality classification under the normal and rule-shift splits,
and malware-detection TPR@0.1\%, where TPR@0.1\% denotes the true-positive rate at a fixed false-positive rate of FPR~=~0.001 (0.1\%). Dashes indicate metrics yet to be reported.}
\label{tab:ablation_pretraining_dataset}
\resizebox{\textwidth}{!}{%
\begin{tabular}{|l|l|cc|cc|cc|c|}
\toprule
\multirow{2}{*}{\textbf{Model}}
& \multirow{2}{*}{\textbf{Pretraining Data}}
& \multicolumn{2}{c|}{\textbf{API call Prediction}}
& \multicolumn{2}{c|}{\textbf{Func. Class. Normal}}
& \multicolumn{2}{c|}{\textbf{Func. Class. RuleShift}}
& \textbf{Malware Detection} \\
\cmidrule{3-9}
& & {\textbf{Top-1} (\%)} & {\textbf{Top-5} (\%)}
& {\textbf{Top-1} (\%)} & {\textbf{Top-3} (\%)}
& {\textbf{Top-1} (\%)} & {\textbf{Top-3} (\%)}
& {\textbf{TPR@0.1\%}} \\
\midrule

\rowcolor{gray!20}
MALPT 4096-B  & Benign PE               & 84.50 & 89.31 & 89.98 & 93.81 & 25.25 & 34.66 & 28.51 \\

MALPT 4096-MB & Malicious \& Benign PE  & 85.87 & 90.27 & \textbf{90.52} & 93.56 & \textbf{25.99} & \textbf{35.67} & 35.89 \\

\rowcolor{gray!20}
MALPT 4096    & Malicious PE            & \textbf{86.51} & \textbf{91.73} & 89.85 &\textbf{94.11} & 25.65 & 35.06 & \textbf{37.84} \\
\bottomrule
\end{tabular}
}
\end{table*}

This appendix reports a pretraining-domain ablation conducted only for the 4,096-token vocabulary setting; its purpose is to isolate the effect of pretraining data composition rather than to compare directly against the main 1,024-token results.

For the development of \sys{}, we trained both the tokenizer and the pretraining model exclusively on the binary code of 35,000 malicious samples.
This design choice was primarily driven by the scarcity of high-quality benign samples and the strategic requirement to reserve our limited pool of benign executables for the testing phase of the malware detection experiment to prevent data leakage.
To rigorously evaluate the impact of this domain-specific pretraining and tokenization, we conducted an ablation study using a newly collected dataset.
We gathered an additional 17,500 benign samples from VirusTotal~\cite{virustotal} that were submitted between March 2025 and June 2025 and that had zero antivirus detections.

Using the BPE 4,096 vocabulary size, we created two distinct pretraining configurations to compare against our baseline: \sys{}-4096-B, trained exclusively on the 17,500 benign samples, and \sys{}-4096-MB, trained on a balanced mixed dataset comprising these 17,500 benign samples and an equal number of 17,500 malicious samples.
To ensure a fair comparison, we trained new tokenizers for each variant to reflect the specific statistical patterns of their respective training corpora.
We then pretrained these models for the exact same number of steps as \sys{}-4096 and fine-tuned them for our downstream tasks using identical hyperparameters.
Table~\ref{tab:ablation_pretraining_dataset} summarizes the available performance results of these configurations.

\mypar{Performance across Domains.}
The available results indicate that all three pretraining configurations achieve strong downstream performance.
For API call prediction, the malware-only model (\sys{}-4096) achieves the highest performance (Top-1: 86.51\%, Top-5: 91.73\%), confirming that malware-specific pretraining aligns the model with the code patterns most relevant for this task.
Functionality classification reveals a more nuanced picture: the mixed-domain model (\sys{}-4096-MB) achieves the strongest accuracy on both the normal and rule-shift splits (Normal Top-1: 90.52\%, Rule-Shift Top-1: 25.99\%), followed by \sys{}-4096-B (Normal Top-1: 89.98\%, Rule-Shift Top-1: 25.25\%) and \sys{}-4096 (Normal Top-1: 89.85\%, Rule-Shift Top-1: 25.65\%).
This suggests that at the 4,096 vocabulary size, exposure to diverse code patterns from both malicious and benign samples during pretraining can benefit functionality recognition.
For malware detection, the malware-only model (\sys{}-4096) achieves the strongest performance (TPR@0.1\% = 37.84\%), followed by the mixed-domain model (\sys{}-4096-MB, TPR@0.1\% = 35.89\%) and the benign-only model (\sys{}-4096-B, TPR@0.1\% = 28.51\%), confirming that malware-specific pretraining provides a consistent advantage for this task.

Although \sys{}-B is also pretrained exclusively on benign binaries, its performance remains strong across all tasks --comparable to \sys{}-- unlike XDA, which performs worse despite being trained on benign data as well (see Section~\ref{sec:eval_comparison_sota}.)
We believe this discrepancy is due to the nature of the benign pretraining corpora used by XDA, and \sys{}-B.
\sys{}-B is trained on benign executables sampled from a broad, real-world distribution, whereas XDA is pretrained on SPEC CPU benchmark binaries, synthetic programs engineered for performance testing rather than representative of real software.
We hypothesize that this selection bias severely limits XDA's exposure to the diverse byte patterns found in real-world executable files during pretraining, leading to substantially weaker downstream performance compared to \sys{}-B.

These results suggest that, while general binary code patterns are shared across domains, malicious binaries contain code patterns that are especially relevant for API call prediction—the task most directly tied to malware behavior.
Malware code often contains both compiler-generated code (similar to benign samples) and non-standard artifacts resulting from obfuscation, packing, or manual assembly.
Consequently, pretraining on malicious code effectively exposes the model to a richer vocabulary of malware-specific code patterns, providing a slight but consistent advantage for API-call-sensitive tasks.
The advantage of mixed-domain pretraining for functionality classification at the 4,096 vocabulary size, on the other hand, suggests that broader code coverage helps the model generalize across a wider range of function behaviors.
Ultimately, these results reinforce our original design choice: pretraining exclusively on malicious code is not merely a constraint-driven compromise, but an effective strategy for maximizing performance in malware-specific applications, particularly where API call modeling matters most.

\begin{takeawaybox}
Across all pretraining data configurations at the 4,096 vocabulary size, the malware-only pretraining gives a slight but consistent boost for API call prediction—the task most tightly coupled with malware-specific code patterns. Functionality classification benefits from broader data diversity, with the mixed-domain model performing best at this vocabulary size. Overall, pretraining on malicious code remains the optimal choice when API call accuracy and malware detection are the primary objectives.
\end{takeawaybox}

\section{Computational Cost}
\label{app:computational_cost}

Automated malware analysis must be timely, but the required throughput varies by deployment context.
Endpoint-level filtering demands millisecond latency, whereas backend analysis pipelines --- where analysts require deeper code understanding --- can tolerate higher per-file cost.
\sys{} is designed for the latter setting.
We report preprocessing, inference, and pretraining costs below; all measurements use a single NVIDIA L40S GPU unless otherwise noted.

\mypar{Preprocessing.}
Table~\ref{tab:preprocessing_cost} reports the time to extract the code section (via \texttt{pefile}) and tokenize it with BPE.
Tokenization cost is flat across vocabulary sizes at approximately 14~ms per file.

\begin{table}[h]
\centering
\caption{Preprocessing time per file (code-section extraction + BPE tokenization) on a single L40S GPU.}
\label{tab:preprocessing_cost}
\begin{tabular}{|l|c|c|}
\toprule
\textbf{Vocab} & \textbf{Mean (ms/file)} & \textbf{Throughput (files/s)} \\
\midrule
512 & 13.7 & 72.8 \\
\rowcolor{gray!20}
1,024 & 13.5 & 74.1 \\
2,048 & 14.1 & 71.0 \\
\bottomrule
\end{tabular}
\end{table}

\mypar{Inference --- Malware Detection.}
Table~\ref{tab:inference_cost} reports per-file inference latency for \sys{}-1024 on the D1 test set (batch size 1).
Latency is dominated by the encoder forward passes; the aggregation head adds negligible overhead ($\sim$2.5~ms median).
The wide gap between median (783~ms) and mean (3,318~ms) reflects the heavy-tailed chunk distribution: median chunk count per binary is 33, but the mean is 152 due to large executables.

\begin{table}[h]
\centering
\caption{Per-file inference latency for malware detection (\sys{}-1024, batch size 1, single L40S GPU).}
\label{tab:inference_cost}
\begin{tabular}{|l|c|c|c|c|}
\toprule
 & \textbf{Total} & \textbf{Encoder} & \textbf{Agg.\ Head} & \textbf{Enc./chunk} \\
\midrule
Mean & 3,318 ms & 3,308 ms & 9.2 ms & 9.29 ms \\
\rowcolor{gray!20}
Median & 783 ms & 780 ms & 2.5 ms & 9.31 ms \\
p90 & 8,908 ms & 8,884 ms & 24.2 ms & 9.37 ms \\
\rowcolor{gray!20}
p99 & 19,068 ms & 19,015 ms & 53.0 ms & 10.77 ms \\
\bottomrule
\end{tabular}
\end{table}

\mypar{Baseline Comparison.}
Table~\ref{tab:baseline_inference_cost} compares per-file inference latency across all malware detection models.
The Ember-based models are the fastest: feature extraction dominates at $\sim$50~ms, with model prediction adding under 2~ms.
Malconv2 is comparable at 54~ms mean.
MalGraph is the slowest model overall, averaging 4.1 minutes per file due to the IDA Pro disassembly required for CFG and FCG construction.
\sys{}-1024 falls between these extremes, slower than the feature-engineering and raw-byte baselines but substantially faster than MalGraph.

\begin{table}[h]
\centering
\caption{Per-file inference latency comparison across malware detection models.}
\label{tab:baseline_inference_cost}
\begin{tabular}{|l|c|c|l|}
\toprule
\textbf{Model} & \textbf{Mean} & \textbf{Median} & \textbf{Notes} \\
\midrule
EmberGBM & 50.7 ms & 10.8 ms & feat.\ extract 49.9 ms + predict 0.8 ms \\
\rowcolor{gray!20}
EmberNN & 52.9 ms & 14.5 ms & feat.\ extract 51.7 ms + forward 1.2 ms \\
Malconv2 & 54.3 ms & 30.5 ms & end-to-end on raw bytes \\
\rowcolor{gray!20}
\sys{}-1024 & 3,318 ms & 783 ms & encoder 9.3 ms/chunk $\times$ $N$ chunks \\
MalGraph & 4.1 min & 22 s & includes IDA Pro disassembly \\
\bottomrule
\end{tabular}
\end{table}

\mypar{Pretraining.}
Table~\ref{tab:pretraining_cost} reports wall-clock time and validation perplexity for each vocabulary configuration, normalized to a budget of 150,000 optimizer steps on 8$\times$L40S GPUs.
Pretraining cost is approximately 81 GPU-hours per configuration regardless of vocabulary size.

\begin{table}[h]
\centering
\caption{Pretraining cost per vocabulary size (150k steps, 8$\times$L40S GPUs).}
\label{tab:pretraining_cost}
\begin{tabular}{|l|c|c|c|}
\toprule
\textbf{Vocab} & \textbf{Steps} & \textbf{Wall-clock (h)} & \textbf{Val.\ PPL} \\
\midrule
256 & 150,000 & 80.9 & 2.30 \\
\rowcolor{gray!20}
512 & 150,000 & 80.3 & 3.72 \\
1,024 & 150,000 & 81.7 & 4.29 \\
\rowcolor{gray!20}
2,048 & 150,000 & 81.9 & 4.95 \\
4,096 & 150,000 & 81.4 & 5.89 \\
\bottomrule
\end{tabular}
\end{table}

\mypar{Fine-Tuning.}
Table~\ref{tab:finetuning_cost} reports the fine-tuning cost for \sys{}-1024 across all three downstream tasks on a single L40S GPU.
Pretrained models (\sys{}, XDA, LMLMs) are fine-tuned for 5 epochs; non-pretrained models (\sys{}-N) and baselines (Malconv2, EmberNN, MLP) are trained for up to 30 epochs.
In all cases, the checkpoint with the lowest validation loss is selected for evaluation.
Malware detection is the most expensive task (1.73 GPU-days) because each sample produces many 1,024-token chunks that must be independently encoded; gradient checkpointing is used to fit training within GPU memory.
API-call prediction and functionality classification operate on single-function sequences and are substantially cheaper.

\begin{table}[h]
\centering
\caption{Fine-tuning cost for \sys{}-1024 (5 epochs, single L40S GPU, batch size 64).}
\label{tab:finetuning_cost}
\begin{tabular}{|l|c|c|c|c|}
\toprule
\textbf{Task} & \textbf{Train samples} & \textbf{Steps} & \textbf{Wall-clock (h)} & \textbf{GPU-days} \\
\midrule
Malware detection & 31,010 & 4,850 & 41.5 & 1.73 \\
\rowcolor{gray!20}
API-call prediction & 453,074 & 35,400 & 4.0 & 0.17 \\
Functionality classif. & 84,918 & 6,635 & 0.75 & 0.03 \\
\bottomrule
\end{tabular}
\end{table}

\section{Malware Detection Error Analysis}
\label{app:malware_detection_error_analysis}

To better characterize the difference between feature-based and code-centric malware detectors, we manually inspected disagreements between EmberGBM and \sys{}-1024.
This analysis is qualitative and is intended only to contextualize the aggregate results in Table~\ref{tab:concept_drift_packed}.
It should not be interpreted as a causal explanation for all errors made by either model.

We first collected malicious and benign executable files that \sys{}-1024 classified correctly but EmberGBM misclassified, either as false positives (benign files detected as malicious) or false negatives (malicious files detected as benign).
For each sample, we used the Ember feature set and extracted the top 20 feature-importance scores assigned by the GBM model.
We analyzed two groups: 20 randomly chosen malicious samples that EmberGBM misclassified as benign, and 20 benign samples that EmberGBM misclassified as malicious.

Across these two groups, EmberGBM's errors were frequently associated with a common set of structural metadata features:  benign samples being greater in size than malicious samples, the presence of a digital signature,  a certificate table (especially of a specific size), a debug directory, and a PE structure with only one read-and-execute memory segment.
When these features were present, EmberGBM tended to classify the file as benign; when absent, it often classified the file as malicious.
This behavior aligns with common properties of legitimate third-party or open-source software and system executables, which are often greater than 1MB in size or often include valid signatures and certificate tables.
However, these fields are not necessarily tied to runtime behavior: an adversary can add randomly generated sections to increase file size, or add a bogus certificate or debug table after compilation without changing program behavior.
This limitation is important because benign datasets are often biased toward publicly available, signed binaries, while benign software in the wild is not always signed~\cite{kaya2025ml}.
Prior work has similarly noted that reliance on signatures is problematic for \emph{certified malware}, where malicious samples carry legitimate code-signing signatures~\cite{kim2017certified,kim2018broken}.

We then inspected the reverse disagreement set: 20 malicious samples that \sys{} misclassified as benign and 20 benign samples that \sys{} misclassified as malicious, while EmberGBM classified them correctly.
Because \sys{} does not expose feature-importance scores, we used static-analysis reports, behavioral-analysis reports, and relational graphs from VirusTotal to characterize these samples.
Among the 20 malicious samples missed by \sys{}, eight were adware, four were packed using custom packers, four contained polymorphic code, three were initial droppers, and one had no available family classification or dynamic-analysis report.
These cases reflect situations where relying only on the code segment can be challenging, either because the malware resembles benign software or because the executable is packed, obfuscated, or encrypted.
In these instances, EmberGBM correctly classified the samples by using non-code structural features such as the certificate table and byte-entropy histogram.

Finally, we inspected the false positives of \sys{}.
All 20 benign samples misclassified by \sys{} exhibited activity categories commonly associated with malware, including privilege escalation, task scheduling, defense evasion, and network communication.
Viewed in isolation, many of these samples are borderline cases.
For 11 out of the 20 samples, \sys{}-1,024's confidence score fell between 0.50 and 0.65, with 0.5 being the classification threshold.
For the same samples, EmberGBM assigned high importance to the presence of a certificate table and debug table.
When we zeroed out the corresponding digital-signature, certificate-table, and debug-table features, EmberGBM's average confidence score on the 20 samples dropped from 0.99 to 0.56; 10 samples fell below the 0.5 threshold, and four fell below 0.1.

Overall, this analysis supports the conclusion that the two model families use complementary evidence.
\sys{} uses code-section bytes and is less exposed to superficial PE metadata, while EmberGBM can benefit from structural features that remain useful for packed, obfuscated, or borderline samples.
This motivates future work on ensembles that combine code-section and PE-structure features, especially under strict low-FPR operating constraints.

\section{Model Architecture and Pretraining Hyperparameters}
\label{app:model_architecture}

Table~\ref{tab:model_hyperparameters} lists the full architecture and pretraining hyperparameters for \sys{}.

\begin{table}[h]
\centering
\caption{Model architecture and pretraining hyperparameters for \sys{}-1024.}
\label{tab:model_hyperparameters}
\begin{tabular}{|l|l|}
\toprule
\textbf{Parameter} & \textbf{Value} \\
\midrule
\multicolumn{2}{|l|}{\textit{Architecture}} \\
\midrule
Base architecture & ModernBERT~\cite{warner2025smarter} \\
\rowcolor{gray!20}
Embedding dim ($d_\text{model}$) & 768 \\
Attention heads ($n_\text{heads}$) & 12 \\
\rowcolor{gray!20}
Transformer layers ($n_\text{layers}$) & 12 \\
FFN inner dim ($d_\text{ff}$) & 3,072 ($4 \times d_\text{model}$) \\
\rowcolor{gray!20}
Total parameters & 86.4M \\
Dropout & 0.1 \\
\rowcolor{gray!20}
Local attention window & 128 tokens \\
Global attention every $N$ layers & 4 \\
\rowcolor{gray!20}
Positional encoding & RoPE ($\theta = 10{,}000$) \\
Normalization & RMSNorm (pre-norm) \\
\rowcolor{gray!20}
Activation & GELU \\
\midrule
\multicolumn{2}{|l|}{\textit{Tokenization}} \\
\midrule
Vocabulary size (v1024) & 1,029 (1,024 BPE + 5 special) \\
\rowcolor{gray!20}
Sequence length & \seqLength{} tokens (\contentLength{} data + \texttt{<s>} + \texttt{</s>}) \\
Pre-tokenizer & ByteLevel, 512-byte windows \\
\midrule
\multicolumn{2}{|l|}{\textit{Pretraining}} \\
\midrule
\rowcolor{gray!20}
Task & Masked Language Modeling (MLM) \\
Loss & Cross-entropy (token-level) \\
\rowcolor{gray!20}
Masking probability & 25\% \\
Masking strategy & 80\% \texttt{<mask>} / 10\% random / 10\% unchanged \\
\rowcolor{gray!20}
Optimizer & Adam ($\beta_1{=}0.9$, $\beta_2{=}0.999$, $\epsilon{=}10^{-6}$) \\
Learning rate & $10^{-4}$ \\
\rowcolor{gray!20}
LR schedule & 1\% linear warmup + polynomial decay \\
Batch size & 128 (64/GPU $\times$ 2 grad.\ accum.) \\
\rowcolor{gray!20}
Precision & BF16 mixed-precision \\
Token budget & \pretrainingTokens{} tokens (150k steps) \\
\rowcolor{gray!20}
GPUs & 8$\times$ NVIDIA L40S \\
\bottomrule
\end{tabular}
\end{table}

\end{document}